\documentclass[10pt,journal]{IEEEtran}
\usepackage{cite}
\usepackage{amsmath,amssymb,amsfonts,amsthm}
\usepackage{color}
\usepackage{algorithmic}
\usepackage{graphicx}
\usepackage{url}
\usepackage{booktabs}
\usepackage[caption=false]{subfig}

\allowdisplaybreaks

\newcommand{\setX}{\mathbf{X}}
\newcommand{\sx}{\mathbf{x}}

\newtheorem{example}{Example}

\newtheorem{theorem}{Theorem}

\newtheorem{lemma}{Lemma}

\begin{document}

\title{Poisson Multi-Bernoulli Approximations for Multiple Extended Object Filtering\\
\author{Yuxuan Xia, Karl Granstr\"{o}m, Lennart Svensson, Maryam Fatemi, {\'A}ngel F. Garc{\'\i}a-Fern{\'a}ndez, Jason L. Williams
\thanks{Y.~Xia, K.~Granstr\"{o}m and L.~Svensson are with the Department of Electrical Engineering, Chalmers University of Technology, G\"{o}teborg, Sweden.
E-mail: firstname.lastname@chalmers.se. M.~Fatemi is with Zenseact, G\"{o}teborg, Sweden. E-mail: maryam.fatemi@zenseact.com. A. F. Garc{\'\i}a-Fern{\'a}ndez is with the Department of Electrical Engineering and Electronics, University of Liverpool, Liverpool, United Kingdom and also with the ARIES research centre, Universidad Antonio de Nebrija, Madrid, Spain. Email: angel.garcia-fernandez@liverpool.ac.uk. J. L. Williams is with the Commonwealth Scientific and Industrial Research Organization, Brisbane, Australia. Email: jason.williams@data61.csiro.au.}}
}

\maketitle

\begin{abstract}
The Poisson multi-Bernoulli mixture (PMBM) is a multi-object conjugate prior for the closed-form Bayes random finite sets filter. The extended object PMBM filter provides a closed-form solution for multiple extended object filtering with standard models. This paper considers computationally lighter alternatives to the extended object PMBM filter by propagating a Poisson multi-Bernoulli (PMB) density through the filtering recursion. A new local hypothesis representation is presented where each measurement creates a new Bernoulli component. This facilitates the developments of methods for efficiently approximating the PMBM posterior density after the update step as a PMB. Based on the new hypothesis representation, two approximation methods are presented: one is based on the track-oriented multi-Bernoulli (MB) approximation, and the other is based on the variational MB approximation via Kullback–Leibler divergence minimisation. The performance of the proposed PMB filters with gamma Gaussian inverse-Wishart implementations are evaluated in a simulation study. 
\end{abstract}


\section{Introduction}
Multi-object tracking (MOT) is the processing of sequences of noisy measurements obtained from multiple sources to estimate the sequences of object states \cite{blackman1999design,challa2011fundamentals,meyer2018message}. Conventional MOT algorithms are usually tailored to the point object assumption: each object is modelled as a point without spatial extent and gives rise to at most one measurement at each time step. However, the point object assumption becomes unrealistic if multiple resolution cells of the sensor are occupied by a single object such that each object may give rise to multiple measurements at each time step. Typical examples of such scenarios include, for example, vehicle and pedestrian tracking using automotive radar and laser range sensors. 

This paper considers the multiple extended object filtering problem where the objective is to estimate both the kinematic and extent states of the current set of objects. An overview of the extended object tracking literature can be found in \cite{extendedoverview}. The detections of an extended object are commonly modelled by an inhomogeneous Poisson point process (PPP), proposed in \cite{ppp}. In the PPP measurement model, the number of detections is Poisson distributed, and the detections are spatially distributed around the object. The PPP measurement model has been integrated into a number of multiple extended object filters including, for example, the extended object joint probabilistic data association (JPDA) filter \cite{hammarstrand2012extended,habtemariam2013multiple,streit2016jpda,vivone2016joint,yang2018linear}, the extended object multiple hypothesis tracker \cite{wieneke2012pmht,coraluppi2018multiple}, filters based on belief propagation \cite{meyer2020scalable,meyer2020scalable2} and several random finite sets (RFSs) based extended object filters \cite{mahler2009phd,phdextended2,phdextended3,cphdextended,lmbextended,pmbmextended2}. It is also possible to model the object detections using a cluster process \cite{swain2010extended}, and it has been integrated into the probability hypothesis density filter \cite{mahler2001detecting,mahler2002extended,swain2010first,swain2011bayesian,swain2012phd,clark2012generalized,swain2013group}. 

RFSs facilitate an elegant Bayesian formulation of the MOT problem, and a significant trend in RFSs-based MOT is the development of multi-object conjugate priors. With PPP birth model, the Poisson multi-Bernoulli mixture (PMBM) is a multi-object conjugate prior for general measurement models \cite{garcia2021poisson}, including, for example, point objects \cite{pmbmpoint} and extended objects with PPP measurement model \cite{pmbmextended2}, in the sense that the PMBM form of the multi-object density is preserved through the filtering recursion. The simulation studies in \cite{pmbmextended,pmbmextended2} have shown that the extended object PMBM filter, in general, performs well in comparison to other RFSs-based extended object filters. 

The PMBM recursion can also handle a multi-Bernoulli (MB) birth model by setting the PPP intensity to zero and adding new Bernoulli components in the prediction step, resulting in the MB mixture (MBM) filter \cite{pmbmpoint2,garcia2019gaussian}. The MBM filter can be further extended to consider MBs with deterministic object existence, referred to as $\text{MBM}_{01}$, at the expanse of an exponential growth in the number of MBs. The labelled $\text{MBM}_{01}$ filtering recursion is analogous to the generalised labelled multi-Bernoulli (GLMB) filter \cite{glmbconjugateprior,lmbextended}. 

The reasoning over the correspondence of measurements and objects, also known as data association, is an inherent challenge in MOT. The unknown data associations lead to an intractably large number of terms in the PMBM density. To achieve computational tractability of the extended object PMBM filter, it is necessary to reduce the number of PMBM parameters, see \cite[Sec. V-B]{pmbmextended2} for different reduction methods. The computational complexity of the PMBM filter, however, still scales with the number of MBs. It is therefore of interest to consider a special variant of the extended object PMBM filter that only propagates a Poisson MB (PMB) density through the filtering recursion.


This paper focuses on developing efficient PMB approximations of the PMBM multi-object posterior density for multiple extended object filtering. We first identify the major challenges in extended object PMB approximation: 1) determining the number Bernoulli components in the approximate PMB; 2) determining which local hypothesis densities should be merged; 3) determining how to merge  the selected local hypothesis densities into one. We then proceed to present a more efficient local hypothesis representation of the PMBM conjugate prior where each measurement creates a new Bernoulli component, whereas for the hypothesis structure in \cite{pmbmextended2,garcia2021poisson} each non-empty subset of measurements creates a new Bernoulli component. This facilitates the development of PMB approximation methods tailored to extended objects. Importantly, the new hypothesis representation results in an equivalent PMBM representation of the multi-object density but with far fewer Bernoulli components.

We present two extended object PMB filters that approximate the extended object PMBM filter using two different PMB approximation methods. The first PMB filter is based on the track-oriented MB (TO-MB) approximation, which was first presented in \cite{pmbmpoint} for point objects and shown to be similar to the JPDA approximation. The idea of using TO-MB approximation for extended object filtering has recently been extended in \cite{garcia2021poisson} to handle scenarios with coexisting point and extended objects, but with a less efficient hypothesis representation where each non-empty subset of measurements creates a new Bernoulli component. The second PMB filter is based on variational approximation that aims to minimise the Kullback–Leibler divergence (KLD) between the PMBM and the approximate PMB. Two different formulations of the minimisation objective are studied: one is based on finding the most likely permutation of local hypotheses for each global hypothesis, and the other follows the linear programming formulation in \cite{variational}. Implementations of the proposed extended object PMB filters, based on gamma Gaussian inverse-Wishart (GGIW) single object models \cite{granstrom2015gamma}, are also presented. Simulation results show that the extended object PMB filters offer an appealing trade-off between computational complexity and estimation performance.

The remainder of the paper is organised as follows. In Section II, we overview the extended object PMBM conjugate prior with standard models and present the problem formulation. The extended object PMB filtering recursion is presented in Section III. The new local hypothesis representation of the PMBM conjugate prior and the TO-MB approximation are presented in Section IV. The variational MB approximation and its two different formulations are presented in Section V. The GGIW implementations of the PMB filters are given in Section VI. The simulation results are shown in Section VII and the conclusions are drawn in Section VIII.

\section{Background and Problem Formulation}
In this section, we first briefly introduce Bayesian MOT and the standard models for multiple extended object tracking. We then overview the extended object PMBM conjugate prior built upon these models. Last, the problem formulation is presented.



\subsection{Bayesian multi-object filtering}

The set of objects at time step $k$ is denoted $\mathbf{X}_k$. By modelling $\mathbf{X}_k$ as a RFS, both object set cardinality $|\mathbf{X}_k|$ and single object state ${\bf x}_k \in {\bf X}_k$ are random variables. In extended object tracking, the object state models both the kinematic properties, and the extent, of the object. The set of measurements at time step $k$ is denoted ${\bf Z}_k = \left\{{\bf z}_k^1,\dots,{\bf z}_k^{m_k}\right\}$, including clutter and object measurements. Further, ${\bf Z}^k = ({\bf Z}_1,\dots,{\bf Z}_k)$ denotes the sequence of measurements from time step $1$ to $k$.

The multi-object filtering density is denoted $f_{k|k}\left({\bf X}_k|{\bf Z}^k\right)$. The multi-object Bayes filter propagates in time the multi-object density $f_{k-1|k-1}\left(\mathbf{X}_{k-1}|\mathbf{Z}^{k-1}\right)$ using the Chapman-Kolmogorov prediction
\begin{multline}
f_{k|k-1}\left(\mathbf{X}_k|\mathbf{Z}^{k-1}\right) \\= \int f_{k,k-1}\left(\mathbf{X}_k|\mathbf{X}_{k-1}\right)f_{k-1|k-1}\left(\mathbf{X}_{k-1}|\mathbf{Z}^{k-1}\right)\delta \mathbf{X}_{k-1},
\label{eq:bayespredict}
\end{multline}
and then updates the density using the Bayes' update
\begin{equation}
f_{k|k}(\mathbf{X}_k|\mathbf{Z}^k) = \frac{f_{k}(\mathbf{Z}_k|\mathbf{X}_k)f_{k|k-1}(\mathbf{X}_k|\mathbf{Z}^{k-1})}{\int f_{k}(\mathbf{Z}_k|\mathbf{X}_k)f_{k|k-1}(\mathbf{X}_k|\mathbf{Z}^{k-1})\delta \mathbf{X}_k},
\label{eq:bayesupdate}
\end{equation}
where $f_{k,k-1}({\bf X}_k|{\bf Z}^{k-1})$ is the multi-object transition density, $f_k({\bf Z}_k|{\bf X}_k)$ is the multi-object measurement set density, and $\int f(\mathbf{X}) \delta \mathbf{X}$ is set integral, defined in \cite[Sec. 11.3.3]{mahler2007statistical}.

\subsection{Multi-object dynamic and measurement models}
\label{section:models}

\subsubsection{Multi-object dynamic model}\label{motionmodel}
Each object survives from time step $k-1$ to time step $k$ with a probability of survival $p^S(\mathbf{x}_{k-1})$. Each survived object evolves independently with a Markovian transition density $\pi_{k,k-1}(\mathbf{x}_{k}|\mathbf{x}_{k-1})$. New objects appear independently of the objects that already exist. The object birth is assumed to be a PPP with intensity $D_{k}^b(\mathbf{x})$. 

\subsubsection{Extended object measurement model}\label{measurementmodel}
The set of measurements $\mathbf{Z}_k$ is a union of a set of clutter measurements and a set of object measurements. The clutter is modelled as a PPP with intensity $\kappa(\mathbf{z})  = \lambda c(\mathbf{z})$, where $\lambda$ is the Poisson clutter rate and $c({\bf z})$ specifies the spatial density of clutter measurements. An extended object with state ${\bf x}_k$ is detected with detection probability $p^D(\mathbf{x}_k)$, and if detected, the object measurements are modelled as a PPP with Poisson rate $\gamma(\mathbf{x}_k)$ and spatial distribution $\phi(\cdot|\mathbf{x}_k)$, independent of any other objects and their corresponding generated measurements.

The conditional extended object measurement set likelihood for a non-empty set of measurements is the product of the probability of detection and the PPP density,
\begin{equation}
    \ell_{\mathbf{Z}}(\mathbf{x}_k) = p^D(\mathbf{x}_k)e^{-\gamma(\mathbf{x}_k)}\prod_{\mathbf{z}\in \mathbf{Z}}\gamma(\mathbf{x}_k)\phi(\mathbf{z}|\mathbf{x}_k).
    \label{eq:extendedlikelihood}
\end{equation}
For an extended object state $\mathbf{x}_k$, the effective probability of detection is the product of the probability of detection and the probability that object generates at least one measurement $1-e^{-\gamma(\mathbf{x}_k)}$. Therefore, the effective probability that the object ${\bf x}_k$ is not detected is
\begin{equation}
    q^D(\mathbf{x}_k) = 1 - p^D(\mathbf{x}_k) + p^D(\mathbf{x}_k)e^{-\gamma(\mathbf{x}_k)}.
    \label{eq:likelihoodemptyset}
\end{equation}
Note that the conditional likelihood for an empty measurement set, $\ell_{\emptyset}(\mathbf{x}_k)$, is also described by (\ref{eq:likelihoodemptyset}).

\subsection{PMBM conjugate prior}

For the above multi-object dynamic and measurement models, the multi-object density $f_{k|k^\prime}(\cdot)$ of ${\bf X}_k$ given the sequence of measurements ${\bf Z}^{k^\prime}$, where $k^\prime \in \{k-1,k\}$, is a PMBM \cite{pmbmextended2}. The PMBM density is of the form \cite{pmbmpoint,xia2019extended}
\begin{subequations}
    \label{eq:pmbm}
    \begin{align}
        f^{\text{pmbm}}_{k|k^\prime}(\mathbf{X}_k) &= \sum_{\mathbf{X}^u_k\biguplus\mathbf{X}^d_k=\mathbf{X}_k}f^{\text{ppp}}_{k|k^\prime}\left(\mathbf{X}^u_k\right)f^{\text{mbm}}_{k|k^\prime}\left({\bf X}^d_k\right),\label{eq:pmbm_explicit}\\
        f^{\text{ppp}}_{k|k^\prime}\left(\mathbf{X}^u_k\right) &= e^{- \int D^u_{k|k^\prime}({\bf x})d {\bf x}}\prod_{\mathbf{x}\in\mathbf{X}_k^u}D^u_{k|k^\prime}(\mathbf{x})\label{eq:ppp},\\
        f^{\text{mbm}}_{k|k^\prime}\left({\bf X}_k^d\right) &= \sum_{a\in{\cal A}_{k|k^\prime}}w^a_{k|k^\prime} \sum_{\biguplus_{l=1}^{n_{k|k^\prime}}{\bf X}^l = {\bf X}_k^d}\prod_{i=1}^{n_{k|k^\prime}}f^{i,a^i}_{k|k^\prime}\left({\bf X}^i\right)\label{eq:mbmdensity},
    \end{align}
\end{subequations}
where $\biguplus$ denotes the disjoint union, $D^u_{k|k^\prime}(\cdot)$ is the intensity of the PPP $f^{\text{ppp}}_{k|k^\prime}(\cdot)$ represent objects that have never been detected (referred to as undetected objects), and $f^{\text{mbm}}_{k|k^\prime}(\cdot)$ is an MBM representing potential objects that have been detected at some point up to time step $k^\prime$. The set of global hypotheses is denoted ${\cal A}_{k|k^\prime}$, and each global hypothesis $a = \left(a^1,\dots,a^{n_{k|k^\prime}}\right)$ is given by selecting a local hypothesis $a^i\in\left\{1,\dots,h^i_{k|k^\prime}\right\}$ for each Bernoulli component $i\in\{1,\dots,n_{k|k^\prime}\}$ subject to certain association constraints \cite{pmbmextended2,xia2019extended}. The global hypothesis weight $w^a_{k|k^\prime}$ is 
\begin{equation}
    \label{eq:globalhypothesisweight}
    w^a_{k|k^\prime} \propto \prod_{i=1}^{n_{k|k^\prime}}w^{i,a^i}_{k|k^\prime},
\end{equation}
where $w^{i,a^i}_{k|k^\prime}$ is the weight of local hypothesis $a^i$, and the proportionality denotes that normalisation is required to ensure that $\sum_{a\in{\cal A}_{k|k^\prime}}w^a_{k|k^\prime} = 1$. In the MBM (\ref{eq:mbmdensity}), there are $n_{k|k^\prime}$ Bernoulli components, and for each Bernoulli there are $h^i_{k|k^\prime}$ possible local hypotheses. The $i$-th Bernoulli with local hypothesis $a^i$ has density
\begin{equation}
    f^{i,a^i}_{k|k^\prime}({\bf X}) = \begin{cases}
        1 - r^{i,a^i}_{k|k^\prime} & {\bf X} = \emptyset,\\
        r^{i,a^i}_{k|k^\prime}p^{i,a^i}_{k|k^\prime}({\bf x}) &{\bf X} = \{{\bf x}\},\\
        0 &\text{otherwise},
    \end{cases}
\end{equation}
where $r^{i,a^i}_{k|k^\prime}$ is the probability of existence and $p^{i,a^i}_{k|k^\prime}(\cdot)$ is the existence-conditioned single object state density.

\subsection{Problem Formulation}

We aim to develop an extended object filter that propagates a PMB density over time, i.e., a special case of the PMBM density (\ref{eq:pmbm}) that has an MBM with a single MB. We denote the PMB predicted/filtering density at time step $k$, with $k^\prime\in\{k-1,k\}$, as 
\begin{equation}
    \label{eq:pmbdensity}
    f^{\text{pmb}}_{k|k^\prime}({\bf X}_k) = \sum_{\biguplus_{l=1}^{n_{k|k^\prime}}{\bf X}^l\biguplus {\bf Y} = {\bf X}_k} f^{\text{ppp}}_{k|k^\prime}({\bf Y})\prod_{i=1}^{n_{k|k^\prime}}f^i_{k|k^\prime}\left({\bf X}^i\right),
\end{equation}
where $f^{\text{ppp}}_{k|k^\prime}(\cdot)$ is of the form (\ref{eq:ppp}) with intensity $D^u_{k|k^\prime}(\cdot)$, $n_{k|k^\prime}$ is the number of Bernoulli components, and the $i$-th Bernoulli component has probability of existence $r^i_{k|k^\prime}$ and existence-conditioned single object state density $p^i_{k|k^\prime}(\cdot)$. 

Given the PMB filtering density $f_{k-1|k-1}^{\text{pmb}}({\bf X}_{k-1})$, the predicted density $f_{k|k-1}^{\text{pmb}}({\bf X}_{k})$ retains the PMB form \cite{pmbmpoint}. The Bayes' update of $f_{k|k-1}^{\text{pmb}}({\bf X}_{k})$, however, results in a PMBM $f_{k|k}^{\text{pmbm}}({\bf X}_{k})$ due to the unknown data associations. To obtain a PMB filtering recursion, the true filtering density $f_{k|k}^{\text{pmbm}}({\bf X}_{k})$ needs to be approximated by a PMB $f_{k|k}^{\text{pmb}}({\bf X}_{k})$, which should retain as much information from $f_{k|k}^{\text{pmbm}}({\bf X}_{k})$ as possible. A natural choice for solving this problem is to minimise the KLD between the PMBM and the approximate PMB, 
\begin{multline}
    \underset{f^{\text{pmb}}}{\min}~D_{\text{KL}}\left(f_{k|k}^{\text{pmbm}}(\mathbf{X}_k) \right\| \left. f^{\text{pmb}}_{k|k}(\mathbf{X}_k)\right) \triangleq \\
    \underset{f^{\text{pmb}}}{\min}~\int f_{k|k}^{\text{pmbm}}(\mathbf{X}_k) \log \left(\frac{f_{k|k}^{\text{pmbm}}(\mathbf{X}_k)}{f^{\text{pmb}}_{k|k}(\mathbf{X}_k)}\right)\delta {\bf X}_k.
    \label{eq:pmbmtopmb}
\end{multline}
Analytically minimising (\ref{eq:pmbmtopmb}) is intractable, and thus we have to use approximations to obtain an efficient algorithm. 

The above problem can be simplified by noting that the PPP representing undetected objects is not hypothesis-dependent, and thus does not require approximation \cite{pmbmpoint}. It is therefore sufficient to consider approximating the MBM $f^{\text{mbm}}_{k|k}\left({\bf X}^d_k\right)$ representing detected objects in $f_{k|k}^{\text{pmbm}}({\bf X}_{k})$ as an MB:
\begin{equation}
    \underset{f^{\text{mb}}}{\arg\min}~D_{\text{KL}}\left(f_{k|k}^{\text{mbm}}\left({\bf X}^d_{k}\right) \right\|\left. f_{k|k}^{\text{mb}}\left({\bf X}^d_{k}\right)\right).
    \label{eq:mbmtomb}
\end{equation}
Lastly, following the recycling method in \cite{recycle}, we note that Bernoulli components with a small probability of existence are well approximated as a PPP. Practically, this allows these Bernoulli components to be modelled through the PPP, reducing the complexity of subsequent data association problems.

\section{Extended Object PMB Filter}
\label{extendedPMB}


In this section, we present the track-oriented extended object PMB filtering recursion, which includes prediction, update, MB approximation and recycling. Compared to the hypothesis-oriented filtering recursion \cite{pmbmextended2}, the track-oriented filtering recursion \cite{xia2019extended} has more efficient hypothesis management and facilitates the development of MB approximation methods.

\subsection{Prediction}

We denote the inner product of $a(x)$ and $b(x)$ as $\langle a;b\rangle\triangleq\int a(x)b(x)dx$ for notational convenience.

\begin{lemma}[PMB prediction]
    \label{lemma_pmbpredict}
    Given the PMB filtering density at time step $k-1$ of the form (\ref{eq:pmbdensity}) and the multi-object dynamic model in Section \ref{motionmodel}, the predicted density at time step $k$ is a PMB of the form (\ref{eq:pmbdensity}), with $n_{k|k-1} = n_{k-1|k-1}$ and
    \begin{subequations}
        \begin{align*}
            D^u_{k|k-1}(\mathbf{x}) &= D^b_k(\mathbf{x})+\left\langle D^u_{k-1|k-1};\pi_{k,k-1}({\bf x}|\cdot)p^S(\cdot)\right\rangle,\\
            r^i_{k|k-1} &= r^i_{k-1|k-1}\left\langle p^i_{k-1|k-1};p^S\right\rangle,\\
            p^i_{k|k-1}(\mathbf{x}) &= \frac{\left\langle p^i_{k-1|k-1};\pi_{k,k-1}({\bf x}|\cdot)p^S(\cdot)\right\rangle}{\left\langle p^i_{k-1|k-1};p^S\right\rangle}.
        \end{align*}
    \end{subequations}
\end{lemma}
Lemma \ref{lemma_pmbpredict} is a special case of the PMBM prediction \cite{pmbmpoint,pmbmextended2} as a PMB is a PMBM with only one mixture component.


\subsection{Update}
We refer to measurement ${\bf z}_k^j$ using the index pair $(k,j)$, and the set of all such index pairs at time step $k$ is denoted ${\cal M}_k$. We also let ${\cal M}_k^{i,a^i}\subseteq {\cal M}_k$ be the set of index pairs associated to local hypothesis $a^i$ of the $i$-th Bernoulli component. In the measurement update step of an extended object PMB filter, every measurement should be assigned to one and only one local hypothesis in a global hypothesis, and more than one measurement can be assigned to the same local hypothesis at the same time step. The set of global hypotheses is then
\begin{multline}
    \label{eq:globalassocconstraint}
    {\cal A}_{k|k} = \Bigg\{ \left(a^1,\dots,a^{n_{k|k}}\right):\\a^i\in\left\{1,\dots,h^i_{k|k}\right\}~\forall~i, \biguplus_{i=1}^{n_{k|k}}{\cal M}_k^{i,a^i} = {\cal M}_k\Bigg\}.
\end{multline}
This means that each global hypothesis, represented as an MB, corresponds to a unique association of measurements ${\bf Z}_k$ to $n_{k|k-1}$ Bernoulli components for previously detected objects and $n_{k|k}-n_{k|k-1}$ new Bernoulli components for newly detected objects. 
\begin{lemma}[PMB update]
    \label{lemma_pmbupdate}
    Given the PMB predicted density at time step $k$ of the form (\ref{eq:pmbdensity}), the multi-object measurement model in Section \ref{measurementmodel}, and a measurement set ${\bf Z}_k = \left\{{\bf z}_k^1,\dots,{\bf z}_k^{m_k}\right\}$, the updated distribution is a PMBM of the form (\ref{eq:pmbm}), with $n_{k|k} = n_{k|k-1} + 2^{m_k} - 1$ and 
    \begin{equation}
        \label{eq:ppp_update}
        D^u_{k|k}({\bf x}) = q^D({\bf x})D^u_{k|k-1}({\bf x}).
    \end{equation}
    For each pre-existing Bernoulli component $f^i_{k|k-1}(\cdot)$, $i\in\{1,\dots,n_{k|k-1}\}$, there are $h^i_{k|k} = 2^{m_k}$ local hypotheses, corresponding to a misdetection and an update using a non-empty subset of ${\bf Z}_k$. The misdetection hypothesis for Bernoulli component $i\in\{1,\dots,n_{k|k-1}\}$ is 
    \begin{subequations}
        \label{eq:mb_miss}
        \begin{align}
            {\cal M}_k^{i,1} &= \emptyset,\label{eq:previous_miss}\\
            w^{i,1}_{k|k} &= 1 - r^i_{k|k-1} + r^i_{k|k-1}\left\langle p^i_{k|k-1};q^D \right\rangle,\\
            r^{i,1}_{k|k} &= \frac{r^i_{k|k-1}\left\langle p^i_{k|k-1};q^D \right\rangle}{1 - r^i_{k|k-1} + r^i_{k|k-1}\left\langle p^i_{k|k-1};q^D \right\rangle},\\
            p^{i,1}_{k|k}({\bf x}) &= \frac{q^D({\bf x})p^{i}_{k|k-1}(\bf x)}{\left\langle p^{i}_{k|k-1};q^D\right\rangle}.
        \end{align}
    \end{subequations}
    
    Let ${\bf Z}_k^1,\dots,{\bf Z}_k^{2^{m_k}-1}$ be the non-empty subsets of ${\bf Z}_k$. The local hypothesis for Bernoulli component $i\in\{1,\dots,n_{k|k-1}\}$ and measurement set ${\bf Z}_k^j$ is
    \begin{subequations}
        \label{eq:mb_update}
        \begin{align}
            {\cal M}_k^{i,j} &= \left\{(k,l):{\bf z}_k^l\in{\bf Z}_k^j\right\},\label{eq:previous_detect}\\
            w^{i,j}_{k|k} &= r^i_{k|k-1}\left\langle p^i_{k|k-1};\ell_{{\bf Z}_k^j} \right\rangle,\\
            r^{i,j}_{k|k} &= 1,\\
            p^{i,j}_{k|k}({\bf x}) &= \frac{\ell_{{\bf Z}_k^j}({\bf x})p^i_{k|k-1}({\bf x})}{\left\langle p^i_{k|k-1};\ell_{{\bf Z}_k^j} \right\rangle}.
        \end{align}
    \end{subequations}
    For the new Bernoulli component ($i=n_{k|k-1}+j$) initiated by measurement set ${\bf Z}_k^j$, there are $h^i_{k|k}=2$ local hypotheses
    \begin{subequations}
        \label{eq:newtracks}
        \begin{align}
            {\cal M}^{i,1}_k &= \emptyset, w_{k|k}^{i,1} = 1, r_{k|k}^{i,1} = 0,\label{eq:nonexistent}\\
            {\cal M}_k^{i,2} &= \left\{(k,l):{\bf z}_k^l\in{\bf Z}_k^j\right\},\label{eq:existent}\\
            w_{k|k}^{i,2} &= \begin{cases}
                \kappa({\bf z}) + \left\langle D^u_{k|k-1};\ell_{{\bf Z}_k^j} \right\rangle & |{\bf Z}_k^j| = 1, \\
                \left\langle D^u_{k|k-1};\ell_{{\bf Z}_k^j} \right\rangle & |{\bf Z}_k^j| > 1,
            \end{cases}\label{eq:new_weight}\\
            r^{i,2}_{k|k} &= \begin{cases}
                \frac{\left\langle D^u_{k|k-1};\ell_{{\bf Z}_k^j} \right\rangle}{\kappa({\bf z}) + \left\langle D^u_{k|k-1};\ell_{{\bf Z}_k^j} \right\rangle} & |{\bf Z}_k^j| = 1, \\ 
                1 & |{\bf Z}_k^j| > 1,
            \end{cases}\label{eq:new_r}\\
            p^{i,2}_{k|k}({\bf x}) &= \frac{\ell_{{\bf Z}_k^j}({\bf x})D^u_{k|k-1}({\bf x})}{\left\langle D^u_{k|k-1};\ell_{{\bf Z}_k^j} \right\rangle}\label{eq:new_p}.
        \end{align}
    \end{subequations}
\end{lemma}
Local hypothesis (\ref{eq:nonexistent}) corresponds to the case that a non-empty subset of ${\bf Z}_k^j$ is associated with another Bernoulli, hence this local hypothesis has probability of existence equal to zero and non-valid single object density. Local hypothesis (\ref{eq:existent}) corresponds to the case that the $i$-th Bernoulli component is created by the set ${\bf Z}_k^j$ of measurements. Lemma \ref{lemma_pmbupdate} is a special case of the PMBM update \cite{pmbmextended2} by considering a PMB prior.

\subsection{MB approximation and its challenges}
\label{section:challenge}

To complete the PMB filtering recursion, we would like to approximate the MBM (\ref{eq:mbmdensity}) with $k^\prime = k$ in the PMBM filtering density
as an MB of the form 
\begin{equation}
    \label{eq:approxmb}
    f^{\text{mb}}_{k|k}({\bf X}_k) = \sum_{\biguplus_{l=1}^{\widetilde{n}_{k|k}}{\bf X}^l = {\bf X}_k}\prod_{i=1}^{\widetilde{n}_{k|k}}\widetilde{f}^i_{k|k}({\bf X}^i).
\end{equation}
There are mainly three challenges in this MB approximation problem. The first challenge is to determine the number of Bernoulli components $\widetilde{n}_{k|k}$ in the approximate MB $f^{\text{mb}}_{k|k}({\bf X}_k)$. Each mixture component in (\ref{eq:mbmdensity}) corresponds to a unique global hypothesis with $n_{k|k}$ local hypotheses, but some of these local hypotheses may correspond to non-existent objects. Considering that $2^{m_k}-1$ new Bernoullis are created at each time step, it may be inefficient to set $\widetilde{n}_{k|k} = n_{k|k-1} + 2^{m_k} - 1$. Since the Bernoulli components in (\ref{eq:mbmdensity}) are order-invariant, the second challenge is to determine which local hypothesis densities $f^{i,a^i}_{k|k}(\cdot)$ ($a\in{\cal A}_{k|k}$, $i\in\{1,\dots,n_{k|k}\}$) should be merged to form $\widetilde{f}^l_{k|k}(\cdot)$ for each $l\in\left\{1,\dots,\widetilde{n}_{k|k}\right\}$. 

\begin{figure*}[!t]
    \centering
    \subfloat[Hypothesis structure]{\includegraphics[width = 1.8\columnwidth]{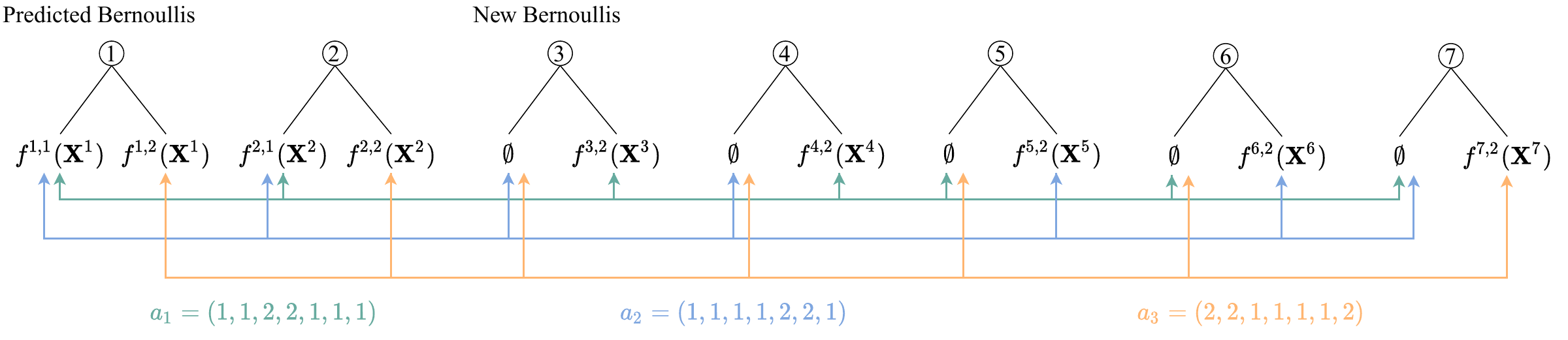}}\\
    \vspace{-2mm}
    \subfloat[Global hypothesis $a_1$]
    {\includegraphics[width = 0.66\columnwidth]{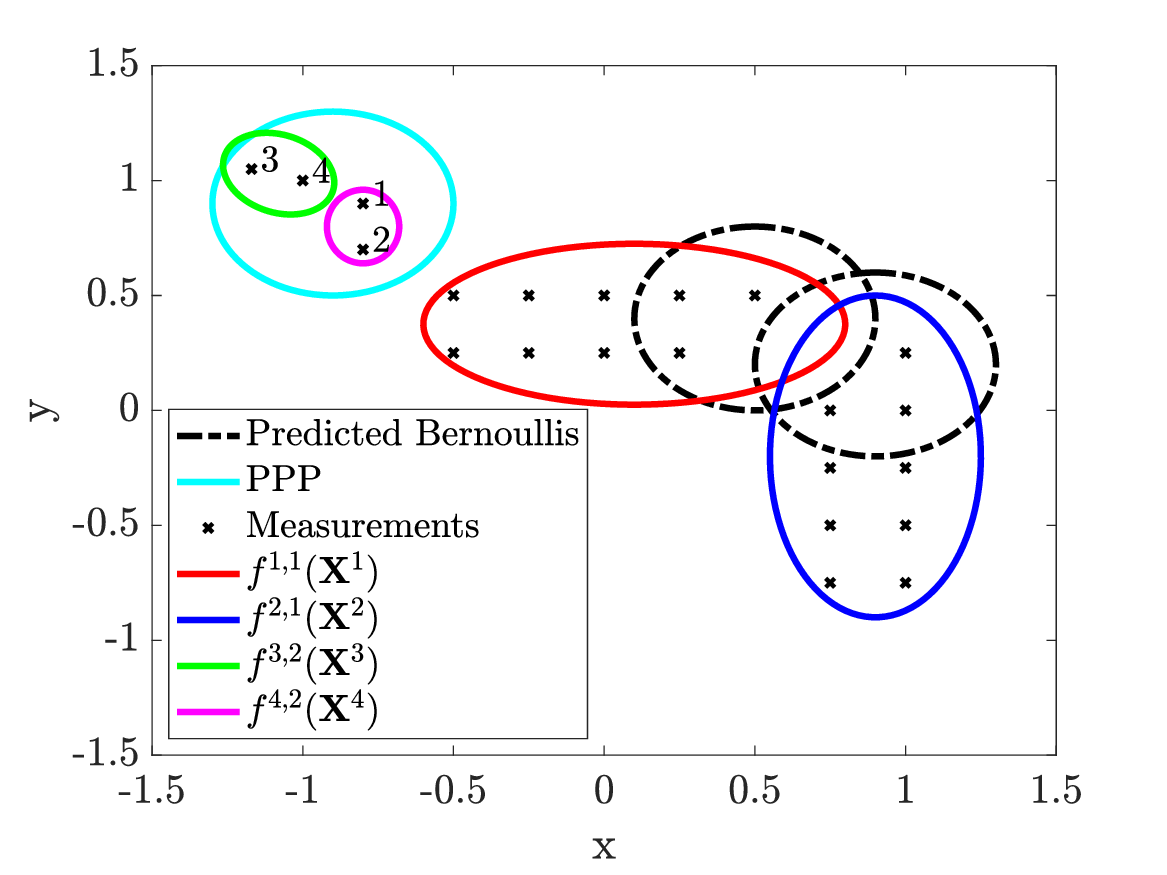}}
    \subfloat[Global hypothesis $a_2$]
    {\includegraphics[width = 0.66\columnwidth]{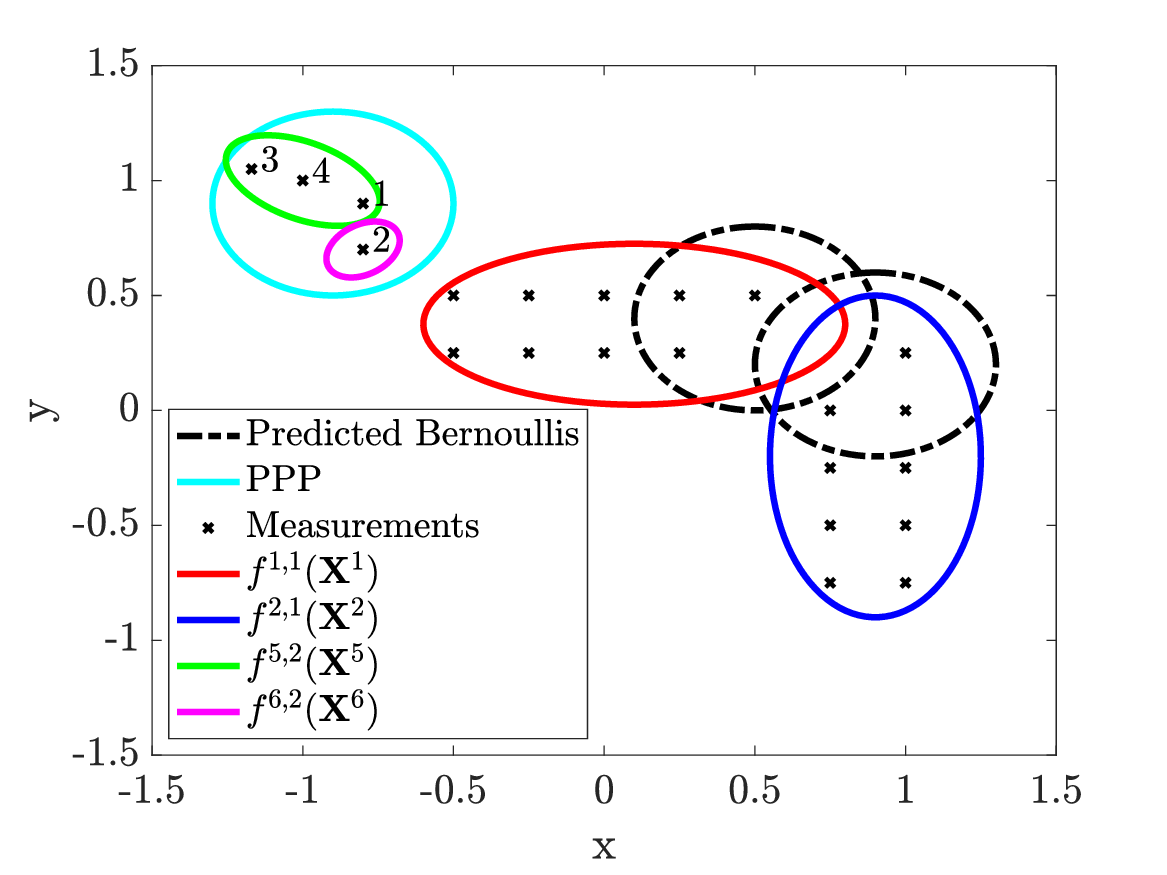}}
    \subfloat[Global hypothesis $a_3$]
    {\includegraphics[width = 0.66\columnwidth]{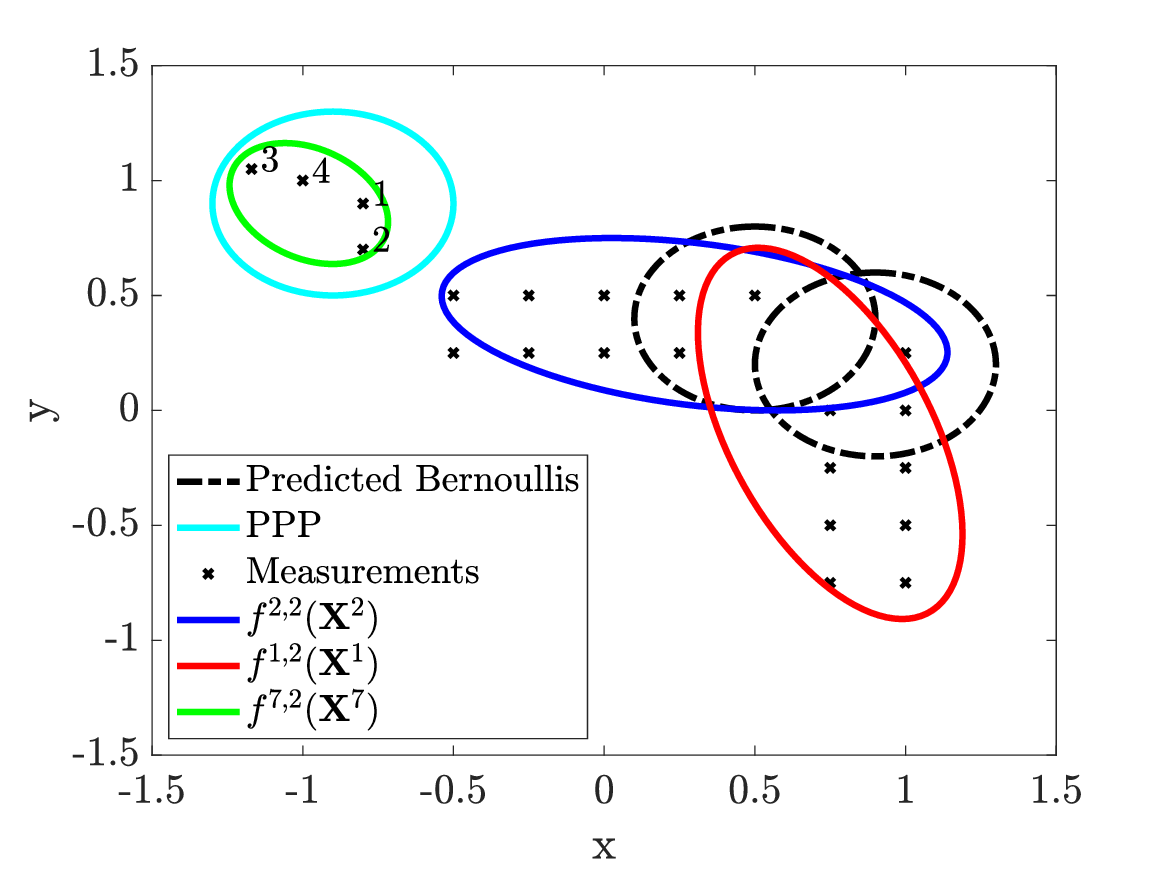}}
    \caption{Illustrative example with three global hypotheses $a_1$, $a_2$ and $a_3$ where $a^i$ denotes the index of the local hypothesis (indexed from left to right) of the $i$-th Bernoulli component. The local hypothesis densities are Gaussian and described by ellipses. Local hypotheses marked in \textcolor{red}{red} and \textcolor{blue}{blue} represent the updates of predicted Bernoullis, whereas local hypotheses marked in \textcolor{green}{green} and \textcolor{magenta}{magenta} represent Bernoullis for the newly detected objects.}
    \label{fig:example}
\end{figure*}

The third challenge is about how to merge the selected local hypothesis densities into one, such that each Bernoulli component in (\ref{eq:mbmdensity}) only corresponds to a single local hypothesis. The single Bernoulli approximation of a Bernoulli mixture can be obtained by taking the weighted sum of different parameters. The approximate Bernoulli component, without further approximation, then contains a single object density collecting mixture components that can be arbitrarily dissimilar. It is usually more desirable to represent the single object density using a unimodal distribution, which simplifies computation and permits closed-form evaluation of objectives, such as the KLD. In extended object filtering, the merging of different single object densities is typically achieved by KLD minimisation, see, e.g., \cite{phdextended,gammareduction}.

We illustrate these challenges with the following example.
\begin{example}
Consider the scenario shown in Fig. \ref{fig:example}, with an MBM representing three global hypotheses. For simplicity, we assume that there is no misdetection, and that each Bernoulli representing an object has probability of existence equal to one. The two objects that are detected for the first time are closely-spaced, so it is difficult to tell whether there are two closely-spaced new objects or one larger new object.

The first challenge is determining how to choose the number of Bernoullis in the approximate MB. In this example, the MBM contains seven Bernoulli components, but each global hypothesis contains local hypotheses corresponding to non-existent objects. Thus, it is non-trivial to select the number of Bernoulli components in the approximate MB.

The second challenge is determining how to select the local hypothesis densities to be merged. For example, when creating approximate Bernoulli components for previously detected objects, we may merge $f^{1,1}(\cdot)$ and $f^{2,2}(\cdot)$ into one, but we may also merge $f^{1,1}(\cdot)$ and $f^{1,2}(\cdot)$ into one. The problem becomes more complicated when we consider how to form approximate Bernoulli components corresponding to the newly detected objects.

The third challenge is determining how to merge the selected local hypothesis densities. For example, if we choose to merge $f^{1,1}(\cdot)$ and $f^{2,2}(\cdot)$, we would like the information loss due to merging to be minimised.
\end{example}

\subsection{Recycling}
\label{section:recycling}
For the approximate MB (\ref{eq:approxmb}), we can apply recycling \cite{recycle} to Bernoulli components with low probability of existence. The recycled Bernoulli components are first approximated as a PPP in the sense of KLD minimisation, and then the approximate PPP is added to the PPP of the PMB filter. After recycling, we set the probability of existence of the recycled Bernoulli components in (\ref{eq:approxmb}) to zero, and the PPP intensity of undetected objects is
\begin{equation*}
    \widetilde{D}_{k|k}^u(\mathbf{x}) = D^u_{k|k}(\mathbf{x}) + \sum_{\substack{i\in \left\{1,\dots,\widetilde{n}_{k|k}\right\}:\widetilde{r}_{k|k}^{i}<\tau_r}}\widetilde{r}^{i}_{k|k}\widetilde{p}_{k|k}^{i}(\mathbf{x}),
\end{equation*}
where $\tau_r$ is the recycling threshold, and $\widetilde{r}^{i}$ and $\widetilde{p}^{i}(\cdot)$ are the probability of existence and the single object density of Bernoulli $\widetilde{f}^i_{k|k}(\cdot)$, respectively. By having fewer Bernoulli components in the MB for detected objects, the complexity of the data association problem in next time step can be reduced. 

\section{Track-Oriented Multi-Bernoulli Approximation}

In this section, we first present a class of possible local hypothesis representations in the PMBM conjugate prior for extended objects. This allows us to find a compact local hypothesis structure for Bernoulli components. Based on an improved local hypothesis representation for the newly detected objects of (\ref{eq:newtracks}), we then present the TO-MB approximation of the MBM (\ref{eq:mbmdensity}) in the PMBM filtering density. We refer to the resulting filter as track-oriented PMB (TO-PMB). The relations of TO-PMB to the labelled MB (LMB) and JPDA for extended objects are also discussed.


\subsection{Local hypothesis structure for newly detected objects}



Let ${\cal F}({\cal M}_k)$ denote the set of all the subsets of ${\cal M}_k$. The following theorem describes a family of hypothesis representations of the updated PMBM posterior that only differ in the representations of the new Bernoulli components; its proof is provided in Appendix \ref{apendix:newlocalhyporepresentation}.

\begin{theorem}
    \label{theorem1}
    The updated PMBM density (\ref{eq:pmbm}), whose set of global hypotheses meets (\ref{eq:globalassocconstraint}), can be equivalently represented by the same local hypotheses for previously detected objects (\ref{eq:mb_miss}), (\ref{eq:mb_update}), and $n_k\geq m_k$ new Bernoulli components with local hypotheses whose associations satisfy
    \begin{subequations}
        \label{eq:constrainsnewlocal}
        \begin{align}
            &\bigcup_{i,j} \left\{{\cal M}_k^{i,j}\right\} = {\cal F}({\cal M}_k),\label{eq:theorem1a}\\ 
            &\forall i~\exists j:{\cal M}_k^{i,j} = \emptyset,\label{eq:theorem1b}\\
            &\forall i,j,j^\prime: j\neq j^\prime\rightarrow{\cal M}_k^{i,j} \neq {\cal M}_k^{i,j^\prime},\label{eq:theorem1c}\\
            &\forall i,i^\prime,j,j^\prime: i\neq i^\prime,j \neq j^\prime, {\cal M}_k^{i,j} \neq \emptyset\rightarrow{\cal M}_k^{i,j} \neq {\cal M}_k^{i^\prime,j^\prime},\label{eq:theorem1d}\\ 
            &\forall i,j,j^\prime:j\neq j^\prime,{\cal M}_k^{i,j}\neq \emptyset,{\cal M}_k^{i,j^\prime}\neq \emptyset\rightarrow{\cal M}_k^{i,j} \cap {\cal M}_k^{i,j^\prime} \neq \emptyset,\label{eq:theorem1e}
        \end{align}
    \end{subequations}
    where the range of variable $i$ is $\{n_{k|k-1}+1,\dots,n_{k|k}\}$ with $n_{k|k} = n_{k|k-1}+n_k$. The density of a new Bernoulli component associated with non-empty ${\cal M}_k^{i,j}$ ($j\in\left\{1,\dots,h^i_{k|k}\right\}$) where ${\cal M}_k^{i,j} = \left\{(k,l):{\bf z}_k^l\in {\bf M}_k^{i,j}\right\}$ is obtained as in (\ref{eq:new_weight})-(\ref{eq:new_p}) by letting ${\bf Z}_k^j = {\bf M}_{k}^{i,j}$.

\end{theorem}
Compared to the representations of the new Bernoulli components (\ref{eq:newtracks}) in Lemma \ref{lemma_pmbupdate} where each non-empty subset of measurements creates a new Bernoulli component, here we consider a more general case where several non-empty subsets may be assigned to the same new Bernoulli component. Constraint (\ref{eq:theorem1a}) is necessary since any non-empty subset of ${\cal M}_k$ can correspond to the first detection of an undetected object. Constraint (\ref{eq:theorem1b}) means that each new Bernoulli component may represent a non-existent object. Constraints (\ref{eq:theorem1c}) and (\ref{eq:theorem1d}) ensure that there do not exist two local hypotheses with the same non-empty subset of ${\cal M}_k$. At last, constraint (\ref{eq:theorem1e}) ensures that two disjoint non-empty subsets of ${\cal M}_k$ cannot be assigned to the same Bernoulli component. This is intuitive as local hypotheses created by measurement clusters with non-empty intersections can never co-exist in a global hypothesis. In other words, for every Bernoulli component $i$, where the range of variable $i$ is $\{n_{k|k-1}+1,\dots,n_{k|k}\}$, there should be at least one measurement which is shared by all the local hypotheses assigned to the component. The constraints (\ref{eq:constrainsnewlocal}) can be interpreted as assigning $2^{m_k}-1$ local hypotheses with non-empty subsets of ${\cal M}_k$ to $n_k\geq m_k$ new Bernoulli components such that local hypotheses of the same Bernoulli component satisfy (\ref{eq:theorem1e}).


\subsection{An alternative PMB update}
\label{sec:alternative_pmb}
In the TO-MB approximation, all the Bernoulli components are forced to be independent, and local hypothesis densities of the same Bernoulli component are merged \cite{pmbmpoint2}. According to Lemma \ref{lemma_pmbupdate}, for a predicted PMB with $n_{k|k-1}$ Bernoulli components and a set ${\bf Z}_k$ of $m_k$ measurements, the TO-MB approximation of the updated distribution contains $\widetilde{n}_{k|k} = n_{k|k-1} + 2^{m_k}-1$ Bernoulli components. However, creating as many new Bernoulli components as there are non-empty measurement subsets may yield intractably many Bernoulli components with low probability of existence. 

Because $m_k$ measurements can correspond to at most $m_k$ newly detected objects, the most efficient class of local hypothesis representations has $n_{k|k} = n_{k|k-1}+m_k$. That is, each measurement creates a new Bernoulli component, as in the case of point object filtering. To construct such a local hypothesis representation that satisfies (\ref{eq:constrainsnewlocal}), one strategy is to choose ${\cal M}_k^{i,1} =\emptyset~\forall~i > n_{k|k-1}$ and $(k,i)\in{\cal M}_k^{i,a^i}~\forall~i > n_{k|k-1},a^i\geq 2$. Let ${\cal M}_k^{i,j-1} = \left\{(k,l):{\bf z}_k^l\in{\bf M}_k^{i,j-1}\right\}~\forall~j\geq 2$ where ${\bf M}_k^{i,j}$ is the $j$-th subset of measurements of the $i$-th Bernoulli component\footnote{Note that a different and varying number of local hypotheses is assigned to each Bernoulli component. This is necessary to correctly represent each association event. For example, if each Bernoulli component was assigned every hypothesis containing the corresponding measurement, many local hypotheses would be duplicated over multiple Bernoulli components, leading to double-counting of association events.}. The set ${\bf S}_i=\left\{ {\bf M}_k^{i,1} ,\dots, {\bf M}_k^{i,i}  \right\}$ of subsets of measurements associated to local hypotheses of the $i$-th Bernoulli component can be built recursively as 
\begin{equation*}
    \label{eq:measurement_build}
    {\bf S}_i = \left\{ \left\{ {\bf z}_k^{i} \right\} \right\} \cup \left( \bigcup_{{\bf M}\in \cup_{j=1}^{i-1}{\bf S}_j} \left\{ \left\{ {\bf z}_k^{i} \right\} \cup {\bf M} \right\}  \right)
\end{equation*}
with ${\bf S}_1 = \left\{ \left\{ {\bf z}_k^{1} \right\} \right\}$. Note that ${\bf S}_i$ is a set of sets and $\left\{ \left\{ {\bf z}_k^{i} \right\} \right\}$ represents a set of single-measurement set $\left\{ {\bf z}_k^{i} \right\}$ \cite[Appendix D]{rfs}. This results in an alternative PMB update, which gives the same PMBM posterior (\ref{eq:pmbm}) as given by the PMB update in Lemma \ref{lemma_pmbupdate}.

\begin{figure}[!t]
    \centering
    \subfloat[Hypothesis structure (Lemma \ref{lemma_pmbupdate})]{\includegraphics[width = \columnwidth]{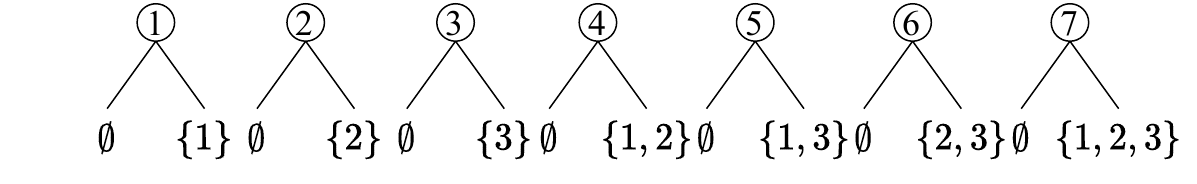}}\\
    \vspace{-2mm}
    \subfloat[Hypothesis structure (Lemma \ref{lemma_newpmbupdate})]
    {\includegraphics[width = \columnwidth]{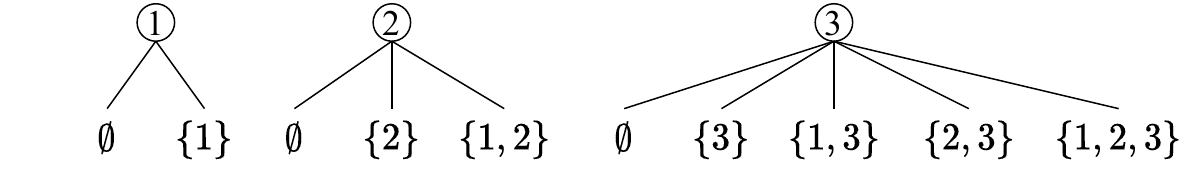}}
    \caption{An illustration of the two hypothesis structures in Example \ref{ex2}, given by Lemma \ref{lemma_pmbupdate} with 7 Bernoulli components (top) and Lemma \ref{lemma_newpmbupdate} with 3 Bernoulli components (bottom), respectively, where the time index $k$ is omitted from the local hypothesis representation. Global hypotheses $a_1$ and $b_1$ correspond to data partition $\{\{1\},\{2\},\{3\}\}$; global hypotheses $a_2$ and $b_2$ correspond to data partition $\{\{1\},\{2,3\}\}$; global hypotheses $a_3$ and $b_3$ correspond to data partition $\{\{1,2\},\{3\}\}$; global hypotheses $a_4$ and $b_4$ correspond to data partition $\{\{1,3\},\{2\}\}$; and global hypotheses $a_5$ and $b_5$ correspond to data partition $\{\{1,2,3\}\}$.}
    \label{fig:example2}
\end{figure}

\begin{lemma}[Alternative PMB update]
    \label{lemma_newpmbupdate}
    Given the PMB predicted density at time step $k$ of the form (\ref{eq:pmbdensity}), the multi-object measurement model in Section \ref{measurementmodel}, and a measurement set ${\bf Z}_k = \left\{{\bf z}_k^1,\dots,{\bf z}_k^{m_k}\right\}$, the updated distribution can be written as a PMBM of the form (\ref{eq:pmbm}), with $n_{k|k} = n_{k|k-1} + m_k$ and updated PPP intensity $D^u_{k|k}({\bf x})$ (\ref{eq:ppp_update}). For each pre-existing Bernoulli component $f^i_{k|k-1}(\cdot)$, $i\in\{1,\dots,n_{k|k-1}\}$, there are $h^i_{k|k} = 2^{m_k}$ local hypotheses, corresponding to a misdetection and an update using a non-empty subset of ${\bf Z}_k$, see (\ref{eq:mb_miss}), (\ref{eq:mb_update}).

    For the new Bernoulli component ($i=n_{k|k-1}+j$), there are $h^i_{k|k}=2^{j-1}+1$ local hypotheses ($\iota\in\left\{1,\dots,2^{j-1}\right\}$)
    \begin{subequations}
        \label{eq:lemma3}
        \begin{align}
            {\cal M}^{i,1}_k &= \emptyset, w_{k|k}^{i,1} = 1, r_{k|k}^{i,1} = 0,\label{eq:nonexistent2}\\
            {\cal M}_k^{i,\iota+1} &= \left\{(k,l):{\bf z}_k^l\in{{\bf M}}_k^{i,\iota}\right\},\label{eq:nonexistent3}\\
            w_{k|k}^{i,\iota+1} &= \begin{cases}
                \kappa({\bf z}) + \left\langle D^u_{k|k-1};\ell_{{{\bf M}}_k^{i,\iota}} \right\rangle & |{\bf M}_k^{i,\iota}| = 1,\\
                \left\langle D^u_{k|k-1};\ell_{{{\bf M}}_k^{i,\iota}} \right\rangle & |{\bf M}_k^{i,\iota}| > 1,
            \end{cases}\\
            r^{i,\iota+1}_{k|k} &= \begin{cases}
                \frac{\left\langle D^u_{k|k-1};\ell_{{{\bf M}}_k^{i,\iota}} \right\rangle}{\kappa({\bf z}) + \left\langle D^u_{k|k-1};\ell_{{{\bf M}}_k^{i,\iota}} \right\rangle} & |{\bf M}_k^{i,\iota}| = 1,\\
                1  & |{\bf M}_k^{i,\iota}| > 1,
            \end{cases}\\
            p^{i,\iota+1}_{k|k}(\bf x) &= \frac{\ell_{{{\bf M}}_k^{i,\iota}}({\bf x})D^u_{k|k-1}({\bf x})}{\left\langle D^u_{k|k-1};\ell_{{{\bf M}}_k^{i,\iota}} \right\rangle}.
        \end{align}
    \end{subequations}
\end{lemma}
For a better understanding of the new hypothesis structure, let us consider the following simple example.
\begin{example}
    \label{ex2}
    Consider the case where $n_{k|k-1} = 0$ and $m_k = 3$. There are 5 different ways to partition three measurements, so the number of global hypotheses is 5. According to Lemma \ref{lemma_pmbupdate}, 7 Bernoulli components are created, and the local hypotheses under these components are: 1) $\emptyset$ and $\{(k,1)\}$, 2) $\emptyset$ and $\{(k,2)\}$, 3) $\emptyset$ and $\{(k,3)\}$, 4) $\emptyset$ and $\{(k,1),(k,2)\}$, 5) $\emptyset$ and $\{(k,1),(k,3)\}$, 6) $\emptyset$ and $\{(k,2),(k,3)\}$, 7) $\emptyset$ and $\{(k,1),(k,2),(k,3)\}$. The 5 global hypotheses represented using these local hypotheses are: $a_1 = (2,2,2,1,1,1,1)$, $a_2 = (2,1,1,1,1,2,1)$, $a_3 = (1,1,2,2,1,1,1)$, $a_4 = (1,2,1,1,2,1,1)$ and $a_5 = (1,1,1,1,1,1,2)$. According to Lemma \ref{lemma_newpmbupdate}, 3 Bernoulli components are created, and the local hypotheses under these components are: 1) $\emptyset$ and $\{(k,1)\}$, 2) $\emptyset$, $\{(k,2)\}$ and $\{(k,1),(k,2)\}$, 3) $\emptyset$, $\{(k,3)\}$, $\{(k,1),(k,3)\}$, $\{(k,2),(k,3)\}$ and $\{(k,1),(k,2),(k,3)\}$. The 5 global hypotheses represented using these local hypotheses corresponding to $a_1$, $a_2$, $a_3$, $a_4$ and $a_5$ are: $b_1=(2,2,2)$, $b_2=(2,1,4)$, $b_3=(1,3,2)$, $b_4=(1,2,3)$ and $b_5=(1,1,5)$, respectively. Note that, compared to the original hypothesis representation, the number of Bernoulli components reduces from 7 to 3, see also Fig. \ref{fig:example2}.
\end{example} 

\subsection{Multi-Bernoulli approximation}
Given the PMBM filtering density resulting from Lemma \ref{lemma_newpmbupdate}, the TO-MB approximation is \cite{pmbmpoint}
\begin{subequations}
    \label{eq:to_mb_approx}
    \begin{align}
        f^{\text{mb}}_{k|k}({\bf X}_k) &= \sum_{\biguplus_{l=1}^{\widetilde{n}_{k|k}}{\bf X}^l = {\bf X}_k}\prod_{i=1}^{\widetilde{n}_{k|k}}\widetilde{f}^i_{k|k}({\bf X}^i),\\
        \widetilde{n}_{k|k} &= n_{k|k-1} + m_k,\\
        w_{k|k}^a &\propto \prod_{i=1}^{\widetilde{n}_{k|k}}w^{i,a^i}_{k|k},\\
        \widetilde{f}^i_{k|k}({\bf X}) &= \sum_{a^i=1}^{h^i}\sum_{b\in{\cal A}_{k|k}:b^i = a^i}w^b_{k|k}f^{i,a^i}_{k|k}({\bf X}).\label{eq:ber_mixture}
    \end{align}
\end{subequations}
The TO-MB approximation (\ref{eq:to_mb_approx}) can be seen to preserve the probability hypothesis density of PMBM filtering density \cite{pmbmpoint2}. It can be also shown that (\ref{eq:to_mb_approx}) minimises the KLD on a single object space augmented with an auxiliary variable\footnote{The KLD on the space of sets of objects with auxiliary variables is an upper bound on the KLD for sets of objects without auxiliary variables \cite{garcia2020trajectory}.}, which indicates if the object remains undetected or corresponds to the $i$-th Bernoulli component \cite{garcia2020trajectory}. Finally, it should be noted that the approximate MB filtering density (\ref{eq:to_mb_approx}) depends on the indices of the individual measurements ${\bf z}_k^j$, with $j\in\{1,\dots,m_k\}$. A different measurement indexing may result in a different approximate MB, but local hypothesis densities created by disjoint measurement sets will never be merged, regardless of the measurement index. As will be discussed in Section \ref{mla}, the problem of finding the local hypothesis representation, that leads to the most accurate TO-MB approximation in the sense of KLD minimisation, can be solved using variational approximation.


\begin{example}
    \label{example3}
    Consider the same scenario and the measurement indexing shown in Fig. \ref{fig:example} and suppose that global hypotheses $a_1$, $a_2$ and $a_3$ have weights $w^{a_1}$, $w^{a_2}$ and $w^{a_3}$, respectively. The PMBM filtering density resulting from Lemma \ref{lemma_newpmbupdate} has two new Bernoulli components: the first one contains local hypotheses $f^{4,2}(\cdot)$ and $f^{6,2}(\cdot)$, whereas the second one contains local hypotheses $f^{3,2}(\cdot)$, $f^{5,2}(\cdot)$ and $f^{7,2}(\cdot)$. The approximate MB obtained using (\ref{eq:to_mb_approx}) contains four Bernoulli components:
    \begin{align*}
        \widetilde{f}^1({\bf X}) &= (w^{a_1}+w^{a_2})f^{1,1}({\bf X}) + w^{a_3}f^{1,2}({\bf X}),\\
        \widetilde{f}^2({\bf X}) &= (w^{a_1}+w^{a_2})f^{2,1}({\bf X}) + w^{a_3}f^{2,2}({\bf X}),\\
        \widetilde{f}^3({\bf X}) &= w^{a_1}f^{4,2}({\bf X}) + w^{a_2}f^{6,2}({\bf X}),\\
        \widetilde{f}^4({\bf X}) &= w^{a_1}f^{3,2}({\bf X}) + w^{a_2}f^{5,2}({\bf X}) + w^{a_3}f^{7,2}({\bf X}).
    \end{align*}
\end{example}

\subsection{Relations to JPDA and LMB}
\label{tomb_filter}
The TO-MB approximation was shown to be exactly the approximation made in the joint integrated probabilistic data association filter \cite{musicki2004joint}, a variant of the JPDA filter by incorporating a probability of existence for each object. That is, the joint association distribution is approximated by the product of its marginals. The main difference is related to the modelling of undetected objects and objects detected for the first time; see \cite{pmbmpoint} for more discussions.

The PMBM filtering recursion can also handle an MB birth model by setting the PPP intensity for undetected objects to zero and adding new Bernoulli components in the prediction step, resulting in the MBM filter \cite{pmbmpoint2,garcia2019gaussian}. Performing the TO-MB approximation after the update step of the MBM filter will then give a TO-MB filter. If the Bernoulli component is uniquely labelled in the MB birth, then the TO-MB filter can be considered as a fast implementation of the LMB filter without $\text{MBM}_{01}$ expansion \cite{olofsson2017sea,kropfreiter2019fast}. Compared to the MB birth model, the Poisson birth model is measurement-driven. The extended object LMB filter can also use an adaptive object birth model that is measurement-driven and allows the newborn objects to appear anywhere in the state space \cite{lmbextended}. However, such adaptive birth model requires careful choice of a few design parameters which are application dependent.

For the LMB filter, object trajectories can be formed by linking object state estimates with the same label. For the PMBM and the TO-PMB filters, object trajectories may be obtained based on filter meta-data (or auxiliary variables \cite{garcia2020trajectory}). For filters based on PMBM conjugate prior, the optimal track building approach is by computing multi-object densities on sets of trajectories \cite{garcia2019multiple}, which has led to, e.g., the trajectory PMBM filter \cite{granstrom2018poisson,xia2019extended} and the trajectory PMB filter \cite{garcia2020trajectory}.

\section{Variational Multi-Bernoulli Approximation}
\label{sec_vmb}


The idea of using variational inference for obtaining an MB approximation that minimises the KLD from the true posterior was first proposed in \cite{variational} for point object filtering. This section presents the variational MB approximation of the MBM in the extended object PMBM posterior density resulted from the PMB update in Lemma \ref{lemma_newpmbupdate} where each measurement creates a new Bernoulli component, i.e., $n_{k|k} = n_{k|k-1}+m_k$. Two different formulations of the minimisation objective in variational approximations are studied: one is based on finding the most likely permutation of local hypotheses in each global hypothesis, and the other follows the linear programming formulation proposed in \cite{variational}.

\subsection{Approximate solution of KLD minimisation}

The objective is to find an MB of the form (\ref{eq:approxmb}) with $n_{k|k} = n_{k|k-1}+m_k$ Bernoulli components that minimises the KLD 
\begin{align}
    &\operatornamewithlimits{argmin}_{\left[\widetilde{f}^i\right]}\int f^{\text{mbm}}_{k|k}({\bf X}_k)\log \left(\frac{f^{\text{mbm}}_{k|k}({\bf X}_k)}{f^{\text{mb}}_{k|k}({\bf X}_k)}\right)\delta {\bf X}_k\nonumber \\ = & \operatornamewithlimits{argmin}_{\left[\widetilde{f}^i\right]} - \int f^{\text{mbm}}_{k|k}({\bf X}_k)\log f^{\text{mb}}_{k|k}({\bf X}_k)\delta {\bf X}_k,
    \label{eq:problem1}
\end{align}
where $f^{\text{mbm}}_{k|k}(\cdot)$ is an MBM of the form (\ref{eq:mbmdensity}), $f^{\text{mb}}_{k|k}(\cdot)$ is an MB of the form (\ref{eq:approxmb}) consisting of Bernoulli components $\widetilde{f}^i_{k|k}(\cdot)$, and $\left[\widetilde{f}^i\right]$ denotes a collection of Bernoulli densities $\widetilde{f}^i_{k|k}(\cdot)$ with $i\in\{1,\dots,n_{k|k}\}$. We further let $\Pi_N$ denote the set of permutation functions on $I_N = \{1,...,N\}$:
\begin{equation*}
    \Pi_N \triangleq \left\{\pi:I_N\rightarrow I_N \mid i\neq j\Rightarrow \pi(i)\neq\pi(j)\right\}.
\end{equation*}
After simplifying the MB set integral in (\ref{eq:problem1}) into a series of Bernoulli set integrals, the solution of (\ref{eq:problem1}) is the same as the solution of the optimisation \cite[Theorem 5]{variational}
\begin{align}
    \label{eq:problem2}
    \operatornamewithlimits{argmin}_{\left[\widetilde{f}^i\right]}&-\sum_{a\in{\cal A}_{k|k}}w^a_{k|k}\int \cdots \int \prod_{i=1}^{n_{k|k}}f^{i,a^i}_{k|k}\left({\bf X}^i\right)\nonumber \\ &\times\log \sum_{\pi\in\Pi_{n_{k|k}}}\prod_{i=1}^{n_{k|k}}\widetilde{f}_{k|k}^{\pi(i)}\left({\bf X}^{\pi(i)}\right)\delta {\bf X}^1\cdots\delta {\bf X}^{n_{k|k}},
\end{align}
which is suitable for applying expectation-maximisation (EM).

Due to the set representation of the multi-object density, we should describe the correspondence between the underlying local hypothesis density $f^{i,a^i}_{k|k}(\cdot)$ in global hypothesis $a$ and the approximate Bernoulli component $\widetilde{f}^i_{k|k}(\cdot)$ using a missing data distribution $q_a(\pi)$, which probabilistically determines which local hypothesis densities in $f^{\text{mbm}}_{k|k}(\cdot)$ should be merged to obtain each approximate Bernoulli component in $f^{\text{mb}}_{k|k}(\cdot)$. Importantly, this relaxes the independence assumption of TO-MB approximation where only local hypothesis densities under the same Bernoulli component can be merged.

By incorporating the missing data distribution $q_a(\pi)$ for each global hypothesis $a$ and using the log-sum inequality, an upper bound of the objective function in (\ref{eq:problem2}) is given by \cite[Sec. III-A]{variational}
\begin{multline}
    \label{eq:problem3}
    J\left(\left[\widetilde{f}^i\right],[q_a(\pi)]\right) = 
    -\sum_{a\in{\cal A}_{k|k},\pi\in\Pi_{n_{k|k}}}w^a_{k|k}q_a(\pi)\\\times\left(-\log q_a(\pi)+\sum_{i=1}^{n_{k|k}}\int f^{i,a^i}_{k|k}\left({\bf X}^i\right)\log \widetilde{f}^{\pi(i)}_{k|k}\left({\bf X}^i\right)\delta {\bf X}^i\right),
\end{multline}
where $[q_a(\pi)]$ denotes a collection of missing data distribution $q_a(\cdot)$ with $a \in {\cal A}_{k|k}$, $q_a(\pi)\geq 0~\forall~a,\pi$ and $\sum_{\pi\in\Pi_{n_{k|k}}}q_a(\pi) = 1~\forall~a$. The optimisation problem (\ref{eq:problem2}) can then be approximately solved by minimising the upper bound (\ref{eq:problem3}):
\begin{equation}
    \label{eq:problem4}
    \min_{\left[\widetilde{f}^i\right],\left[q_a(\pi)\right]}J\left(\left[\widetilde{f}^i\right],[q_a(\pi)]\right).
\end{equation}
An approximate solution of (\ref{eq:problem4}) can be obtained using EM by coordinate descent\footnote{In most applications of EM, the missing data distribution is estimated for each finite number of training samples. In turn, this guarantees that the log-sum inequality is tight at the optimum, and hence that the procedure will converge to a local minimum of (\ref{eq:problem4}) \cite{variational}.}, which proceeds by alternating between estimating the missing data distribution $q_a(\pi)$ with $a\in{\cal A}_{k|k}$ (E-step), and optimising $\widetilde{f}^i_{k|k}(\cdot)$ with $i\in\{1,\dots,n_{k|k}\}$ that minimises (\ref{eq:problem3}) (M-step). These two steps can be solved as 
\begin{align}
    q_a(\pi) &\propto \prod_{i=1}^{n_{k|k}}\exp \left( \int f^{i,a^i}_{k|k}({\bf X})\log \widetilde{f}^{\pi(i)}_{k|k}({\bf X})\delta {\bf X} \right),\label{eq:estep}\\
    \widetilde{f}_{k|k}^i({\bf X}) &= \sum_{a\in{\cal A}_{k|k},\pi\in\Pi_{n_{k|k}}}w^aq_a(\pi)f_{k|k}^{\pi^{-1}(i),a^{\pi^{-1}(i)}}({\bf X}),\label{eq:mstep}
\end{align}
where $\pi^{-1}(\cdot)$ is the inverse of the permutation function $\pi(\cdot)$. 

\subsection{Finding the most likely assignment}
\label{mla}

Exact solution of the missing data distribution $q_a(\pi)$ in (\ref{eq:estep}) may be computed by enumerating all the possible permutations of local hypothesis densities contained in global hypothesis $a$ in a brute force fashion. Approximate solutions of (\ref{eq:estep}) may also be obtained by only considering permutations with high weights using e.g., Murty's algorithm \cite{miller1997optimizing} or Gibbs sampling \cite{fatemi2017poisson}. Empirically, we have found that it is more efficient from a computational perspective to only consider the most likely permutation of local hypothesis densities for each global hypothesis. In this case, the missing data distribution under each global hypothesis becomes a point mass, and the E-step (\ref{eq:estep}) and the M-step (\ref{eq:mstep}) reduce to 
\begin{align}
    \label{eq:ml_assign}
    \operatornamewithlimits{argmin}_{\widetilde{\pi}_a}&-\sum_{i=1}^{n_{k|k}}\int f^{i,a^i}_{k|k}({\bf X})\log \widetilde{f}^{\widetilde{\pi}_a(i)}_{k|k}({\bf X})\delta {\bf X},\\
    \widetilde{f}_{k|k}^i({\bf X}) &= \sum_{a\in{\cal A}_{k|k}}w^af_{k|k}^{\widetilde{\pi}_a^{-1}(i),a^{\widetilde{\pi}_a^{-1}(i)}}({\bf X}),\label{eq:ml_mstep}
\end{align}
where the most likely assignment $\widetilde{\pi}_a$ for each global hypothesis $a\in{\cal A}_{k|k}$ can be obtained by solving a 2D assignment problem with computational complexity ${\cal O}\left(n_{k|k}^3\right)$ \cite{crouse2016implementing}. Hence, the total computational complexity of solving (\ref{eq:ml_assign}) for all the global hypotheses is ${\cal O}\left(|{\cal A}_{k|k}|n_{k|k}^3\right)$. In practical implementations, different approximation methods can be used to keep the number of global hypotheses $|{\cal A}_{k|k}|$ at a tractable level; this is discussed in Section \ref{sec:dataassoc}.

Because the MB distribution is invariant to the indices of the Bernoulli components it contains, the selection of the permutation function $\widetilde{\pi}_a$ for each global hypothesis $a\in{\cal A}_{k|k}$ will not change the MBM $f^{\text{mbm}}_{k|k}(\cdot)$, but it determines which local hypothesis densities should be merged. In TO-MB approximation, local hypothesis densities $f^{i,a^i}_{k|k}(\cdot)$ corresponding to the same Bernoulli component, i.e., with the same superscript $i$, are merged. The simplified E-step (\ref{eq:ml_assign}) can be understood as constructing an alternative local hypothesis representation where local hypothesis densities $f_{k|k}^{\widetilde{\pi}_a^{-1}(i),a^{\widetilde{\pi}_a^{-1}(i)}}(\cdot)$ with $a\in{\cal A}_{k|k}$ correspond to the $i$-th Bernoulli component, whereas the simplified M-step (\ref{eq:ml_mstep}) is equivalent to applying TO-MB approximation to $f^{\text{mbm}}_{k|k}(\cdot)$ using the alternative local hypothesis representation determined by (\ref{eq:ml_assign}). This closely connects to the problem of searching for the local hypothesis representation for newly detected objects that yields the most accurate TO-MB approximation in the sense of KLD minimisation; the difference here is that the optimisation problem is now jointly solved for all the detected objects.


It was also shown in \cite{variational} that the missing data distribution in the above formulation acts in an equivalent manner to the selection of an ordered distribution in the same unordered family that can be approximated by the desired distribution family in the Kullback–Leibler JPDA filter \cite{sjpda}. However, note that the related optimisation problem in \cite{sjpda} is solved using importance sampling, which is more computationally demanding than solving 2D assignment problems.

\subsection{Efficient approximation of the feasible set}
\label{eafs}

It was observed in \cite{variational} that the EM algorithm with the objective function 
\begin{multline}
    \label{eq:problem5}
    \widetilde{J}\left(\left[\widetilde{f}^i\right],[q_a(\pi)]\right) = 
    -\sum_{a\in{\cal A}_{k|k},\pi\in\Pi_{n_{k|k}}}w^a_{k|k}q_a(\pi)\\\times\sum_{i=1}^{n_{k|k}}\int f^{i,a^i}_{k|k}\left({\bf X}^i\right)\log \widetilde{f}^{\pi(i)}_{k|k}\left({\bf X}^i\right)\delta {\bf X}^i
\end{multline}
tends to yield solution with lower KLD. For notational convenience, let the local hypotheses $f^{h}_{k|k}(\cdot)$ for all the Bernoulli components be indexed through the set ${\cal H}_{k|k}$ with $|{\cal H}_{k|k}| = \sum_{i=1}^{n_{k|k}}h^i_{k|k}$. The minimisation of (\ref{eq:problem5}) can be solved equivalently as \cite[Theorem 6]{variational}
\begin{equation}
    \label{eq:problem6}
    \min_{q(h,j)\in{\cal P}}-\sum_{j=1}^{n_{k|k}}\int \left( \sum_{h\in{\cal H}_{k|k}}q(h,j)f^h_{k|k}({\bf X}) \right) \log \widetilde{f}_{k|k}^j({\bf X}) \delta {\bf X},
\end{equation}
where the polytope ${\cal P}$ is 
\begin{multline}
    \label{eq:polytope_P}
    {\cal P} = \left\{ q(h,j) = \sum_{i=1}^{n_{k|k}} \sum_{a\in{\cal A}_{k|k}|a^i=h}w^a \sum_{\pi\in\Pi_{n_{k|k}}|\pi(i)=j}q_a(\pi) \right. \\ \left | q_a(\pi)\geq 0, \sum_{\pi\in\Pi_{n_{k|k}}}q_a(\pi)=1 \right\},
\end{multline}
$a = \left(a^1,\dots,a^{n_{k|k}}\right)$ and $q(h,j)$ specifies the weight of local hypothesis $f^{h}_{k|k}(\cdot)$ in Bernoulli component $\widetilde{f}_{k|k}^j(\cdot)$. 

An approximate solution of (\ref{eq:problem6}) can be obtained by considering a relaxation of the polytope ${\cal P}$ 
\begin{multline}
    \label{eq:polytope_M}
    {\cal M} = \left\{ q(h,j)\geq 0 \left|  \sum_{h\in{\cal H}_{k|k}}q(h,j) = 1,\right.\right.\\ \left.\sum_{j=1}^{n_{k|k}}q(h,j) = \sum_{i=1}^{n_{k|k}}\sum_{a\in{\cal A}_{k|k}|a^i=h}w^a\right\}.
\end{multline}
Compared to (\ref{eq:polytope_P}), the missing data distribution in (\ref{eq:polytope_M}) is not constrained to depend on $q_a(\pi)$. In this case, the E-step (\ref{eq:estep}) and the M-step (\ref{eq:mstep}) reduce to 
\begin{align}
    \label{eq:tp_lp}
    \operatornamewithlimits{argmin}_{\widetilde{q}(h,j)\in{\cal M}} &-\sum_{h\in{\cal H}_{k|k}}\sum_{j=1}^{n_{k|k}}\widetilde{q}(h,j)\int f^{h}_{k|k}({\bf X})\log \widetilde{f}^j_{k|k}({\bf X})\delta {\bf X},\\
    \widetilde{f}^j_{k|k}({\bf X}) &= \sum_{h\in{\cal H}_{k|k}}\widetilde{q}(h,j)f^h_{k|k}({\bf X}),\label{eq:tp_mstep}
\end{align}
where the optimisation problem involved in (\ref{eq:tp_lp}) is equivalent to the minimum-cost flow problem that can be solved by linear programming in polynomial time \cite[Chap. 7]{papadimitriou1998combinatorial}. The computational complexity of solving (\ref{eq:tp_lp}) depends on the number of Bernoulli components $n_{k|k}$ and the number of local hypotheses ${\cal H}_{k|k}$, but not the number of global hypotheses $|{\cal A}_{k|k}|$, as opposed to the optimisation problem in (\ref{eq:ml_assign}).

\subsection{An illustrative example}

The variational MB approximation involves an iterative update of the missing data distribution and the approximate MB. When initialised using the marginal association probabilities, the variational MB approximation may be considered an improvement on the TO-MB approximation in the sense that an upper bound of the KLD (\ref{eq:problem1}) between the MBM and the approximate MB is minimised. We illustrate how the missing data distributions change from one iteration of the variational MB approximation to the next for the two different optimisation methods, respectively, in the following example.



\begin{example}
Consider the same scenario shown in Fig. \ref{fig:example}. Suppose that we have used the TO-MB approximation to obtain an approximate MB with four Bernoulli components, as described in Example \ref{example3}.

For the variational MB approximation using 2D assignment, we can find the most likely permutation of Bernoulli components for each MB according to (\ref{eq:ml_assign}). In this case, 
the sum of the cross entropy between $f^{2,2}(\cdot)$ and $\widetilde{f}^1(\cdot)$ and the cross entropy between $f^{1,2}(\cdot)$ and $\widetilde{f}^2(\cdot)$ (cf. (\ref{eq:ml_assign})) is smaller than the sum of the cross entropy between $f^{1,2}(\cdot)$ and $\widetilde{f}^1(\cdot)$ and the cross entropy between $f^{2,2}(\cdot)$ and $\widetilde{f}^2(\cdot)$. In order to minimise the objective (\ref{eq:problem4}), the order of the first two Bernoulli components in the third MB should be swapped. After the reordering, the first two approximate Bernoulli components is 
\begin{align*}
    \widetilde{f}^1({\bf X}) &= (w^{a_1}+w^{a_2})f^{1,1}({\bf X}) + w^{a_3}f^{2,2}({\bf X}),\\
    \widetilde{f}^2({\bf X}) &= (w^{a_1}+w^{a_2})f^{2,1}({\bf X}) + w^{a_3}f^{1,2}({\bf X}).
\end{align*}

Under the same assumption, for the variational MB approximation using linear programming, in order to minimise the objective (\ref{eq:problem5}), a proportion of the weight of assigning $f^{1,2}(\mathbf{X})$ to $\widetilde{f}^1(\mathbf{X})$ should be shifted to $\widetilde{f}^2(\mathbf{X})$, and accordingly a proportion of the weight of assigning $f^{2,2}(\mathbf{X})$ to $\widetilde{f}^2(\mathbf{X})$ should be shifted to $\widetilde{f}^1(\mathbf{X})$. For example, the first two approximate Bernoulli components after weight shifting may become
\begin{multline*}
    \widetilde{f}^1({\bf X}) = (w^{a_1}+w^{a_2})f^{1,1}({\bf X}) \\+ (w^{a_3}-\Delta w)f^{2,2}({\bf X}) + \Delta w f^{1,2}({\bf X}),
\end{multline*}
\begin{multline*}
    \widetilde{f}^2({\bf X}) = (w^{a_1}+w^{a_2})f^{2,1}({\bf X}) \\+ (w^{a_3} - \Delta w)f^{1,2}({\bf X}) + \Delta w f^{2,2}({\bf X}),
\end{multline*}
where both the weight of $f^{1,2}({\bf X})$ in $\widetilde{f}^1({\bf X})$ and the weight of $f^{2,2}({\bf X})$ in $\widetilde{f}^2({\bf X})$ are $0 \leq \Delta w \leq w^{a_3}$, given by the optimal $\tilde{q}(\cdot,\cdot)$ via solving (\ref{eq:tp_lp}).
\end{example}

\section{GGIW Implementation}
Solving the multiple extended object filtering problem requires not only an MOT filtering recursion, but also a single extended object model. There are several single extended object models available in the literature, including, for example, the random matrix model \cite{randomMatrix,randomMatrix2,tuncer2021random}, the multiplicative error model \cite{yang2019tracking}, the random hypersurface model \cite{baum2014extended} and its variant Gaussian process model \cite{wahlstrom2015extended}. In this paper, we have chosen the random matrix model in \cite{randomMatrix2}, in which the object shape is approximated as an ellipse. The random matrix model is simple to use, yet flexible enough to be applicable to many real scenarios, e.g., pedestrian tracking using laser scanner \cite{phdextended3,soextended} and ship tracking using marine radar \cite{schuster2015probabilistic,granstrom2015gamma}. The random matrix model has also been used in several multiple extended object filters, see, e.g., \cite{phdextended3,cphdextended,lmbextended,pmbmextended2}.




\subsection{Single object model}
In the random matrix model, the extended object state $\mathbf{x}_k$ is the combination of a vector $\xi_k$ describing the object kinematic state, a matrix $\chi_k$ describing the object extent and a scalar $\gamma_k$ being the measurement model Poisson rate.

The measurement likelihood for a single measurement $\mathbf{z}$ is
\begin{equation}
    \phi(\mathbf{z}_k|\mathbf{x}_k) = \mathcal{N}(\mathbf{z}_k;H\xi_k,z\chi_k+R_k),
    \label{eq:mealikelihood}
\end{equation}
where $H$ selects the object position from $\xi_k$, $z$ is a scaling factor, and $R_k$ is the covariance of the Gaussian measurement noise. The single object density for the PPP model (\ref{eq:ppp}) with single measurement
likelihood (\ref{eq:mealikelihood}) is a product of gamma, Gaussian and inverse-Wishart distributions \cite{randomMatrix2,granstrom2015gamma}
\begin{subequations}
    \begin{align}
        p(\mathbf{x}) &= \mathcal{GAM}(\gamma;a,b)\mathcal{N}(\xi;m,P)\mathcal{IW}_d(\chi;v,V),\label{eq:ggiw}\\ &\triangleq \mathcal{GGIW}(\mathbf{x};\zeta),
        \end{align}
\end{subequations}
where $\mathcal{GAM}(\gamma;a,b)$ is the gamma distribution defined on $\gamma > 0$ with shape $a$ and rate $b$, $\mathcal{IW}_d(\chi;v,V)$ is the inverse-Wishart distribution defined on positive definite $d\times d$ matrix with degrees of freedom $v$ and $d\times d$ scale matrix $V$, and $\zeta = (a,b,m,P,v,V)$ is the GGIW density parameters.

The motion models are given by
\begin{subequations}
    \label{eq:motionmodel}
\begin{align}
    \xi_{k} &= g(\xi_{k-1}) + w_{k-1},\\
    \chi_{k} &= M(\xi_{k-1})\chi_{k-1}M(\xi_{k-1})^T,\\
    \gamma_{k} &= \gamma_{k-1},
\end{align}
\end{subequations}
where $g(\cdot)$ is a kinematic motion model, $w_{k-1}$ is Gaussian process noise with zero mean and covariance $Q$, and $M(\cdot)$ is a transformation matrix. 

\subsection{Bernoulli-GGIW distribution}


To solve the optimisation problems (\ref{eq:ml_assign}) and (\ref{eq:tp_lp}), the cross entropy between two Bernoulli-GGIW distributions needs to be computed. Because the gamma distribution, the Gaussian distribution and the inverse-Wishart distribution in (\ref{eq:ggiw}) are mutually independent, analytical expression of such cross entropy exists; see Appendix \ref{appendix:crossentropy} for details.

To approximate (\ref{eq:ber_mixture}), (\ref{eq:ml_mstep}) and (\ref{eq:tp_mstep}) as a single Bernoulli component, we also need expressions for merging a Bernoulli-GGIW mixture distribution. Suppose that we have a mixture of Bernoulli components indexed by $h\in\mathcal{H}$ and that the $h$-th Bernoulli component has weight $w^h$, probability of existence $r^h$ and GGIW density $\mathcal{GGIW}(\mathbf{x}^h;\zeta^h)$. According to the result in Appendix \ref{appendix:bernoullimixturereduction}, the best-fitting Bernoulli-GGIW in the sense of KLD minimisation has probability of existence
\begin{equation*}
    \widetilde{r} = \sum_{h\in\mathcal{H}} w^hr^h,
\end{equation*}
and GGIW parameters
\begin{equation*}
    \operatornamewithlimits{argmin}_{\widetilde{\zeta}}~D_{\text{KL}} \left.\left(  \frac{\sum_{h\in\mathcal{H}} w^{h} r^{h} \mathcal{GGIW}(\mathbf{x}^h;\zeta^h)}{\sum_{h\in\mathcal{H}} w^{h}r^{h}}  \right\| \mathcal{GGIW}(\mathbf{x};\widetilde{\zeta})\right).
\end{equation*}
The merging of multivariate Gaussian mixture that minimises the KLD can be achieved using moment matching. The numerical methods for merging gamma mixture and inverse-Wishart mixture via analytical KLD minimisation are presented in \cite{phdextended} and \cite{gammareduction}, respectively.

\subsection{Approximations for computational tractability}
\label{sec:dataassoc}
The pseudo code for the prediction and the update of PPP-GGIW and Bernoulli-GGIW can be found in \cite{pmbmextended2}, so we focus on approximation methods for keeping the computational complexity of the GGIW-PMB filter at a tractable level.

First, gating is performed to remove unlikely measurement-to-object associations. In extended object filtering, the gates need to take into account both the position and extent of the object, as well as the state uncertainties. Second, clustering algorithms with different hyperparameter settings are used to obtain several different partitions of the measurements. Last, for each measurement partition we only consider  cluster-to-object assignments with relatively high likelihoods, which can be obtained using, e.g., ranked assignment \cite{miller1997optimizing} or Gibbs sampling \cite{fatemi2017poisson}. The above clustering and assignment procedure usually results in a small subset of the cluster-to-object assignments with relatively high likelihoods, and therefore greatly reduces the number of possible data associations in the update step. Alternative approaches to handling the data association problem in extended object filtering include the random sampling methods \cite{soextended} and the sum-product algorithms \cite{meyer2020scalable,meyer2020scalable2}.

When performing MB approximation, it is computationally lighter to only consider MBs with non-negligible normalised weight. Thus, we prune MBs whose updated weight fall below a threshold before MBM merging. After MB approximation, Bernoulli components with probability of existence below a threshold are recycled and added to the updated PPP intensity representing undetected objects, see Section \ref{section:recycling}. 
The PPP intensity is typically represented by a mixture. To reduce the computational complexity of the measurement update of PPP, mixture components with weight below a threshold are pruned from the PPP intensity, and similar mixture components, in the sense of small KLD, can be merged. We emphasise that the approximations are applied to the elements $a$ of the outer sum in (\ref{eq:mbmdensity}) the elements of $D^u(\cdot)$ in (\ref{eq:ppp}). The summations mapping elements of ${\bf X}_k$ on the LHS of (\ref{eq:pmbm_explicit}) to ${\bf X}_k^u$ and ${\bf X}_k^d$, and mapping the elements of ${\bf X}_k^d$ on the LHS to ${\bf X}_k^1,\dots,{\bf X}_k^n$ ensure symmetry of the respective distributions regardless of the pruning and recycling approximations that we apply. These symmetrising summations do not need to be explicitly evaluated in prediction, update or estimation; they remain implicit throughout and are not subject to approximation.

\section{Simulations and Results}

\begin{figure*}[t!]
    \centering
    \subfloat[Scenario with 96 objects]{\includegraphics[width = 0.67\columnwidth]{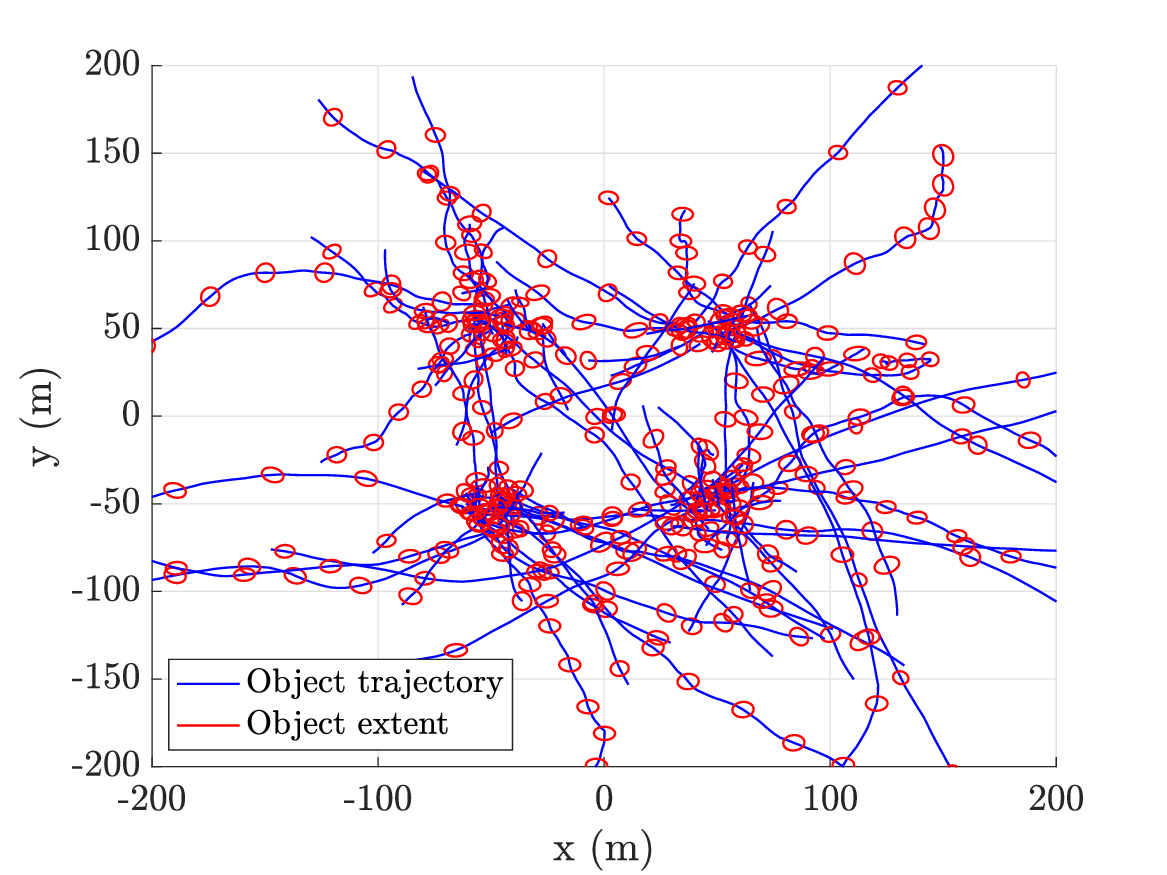}}
    \subfloat[Scenario with 26 objects]
    {\includegraphics[width = 0.67\columnwidth]{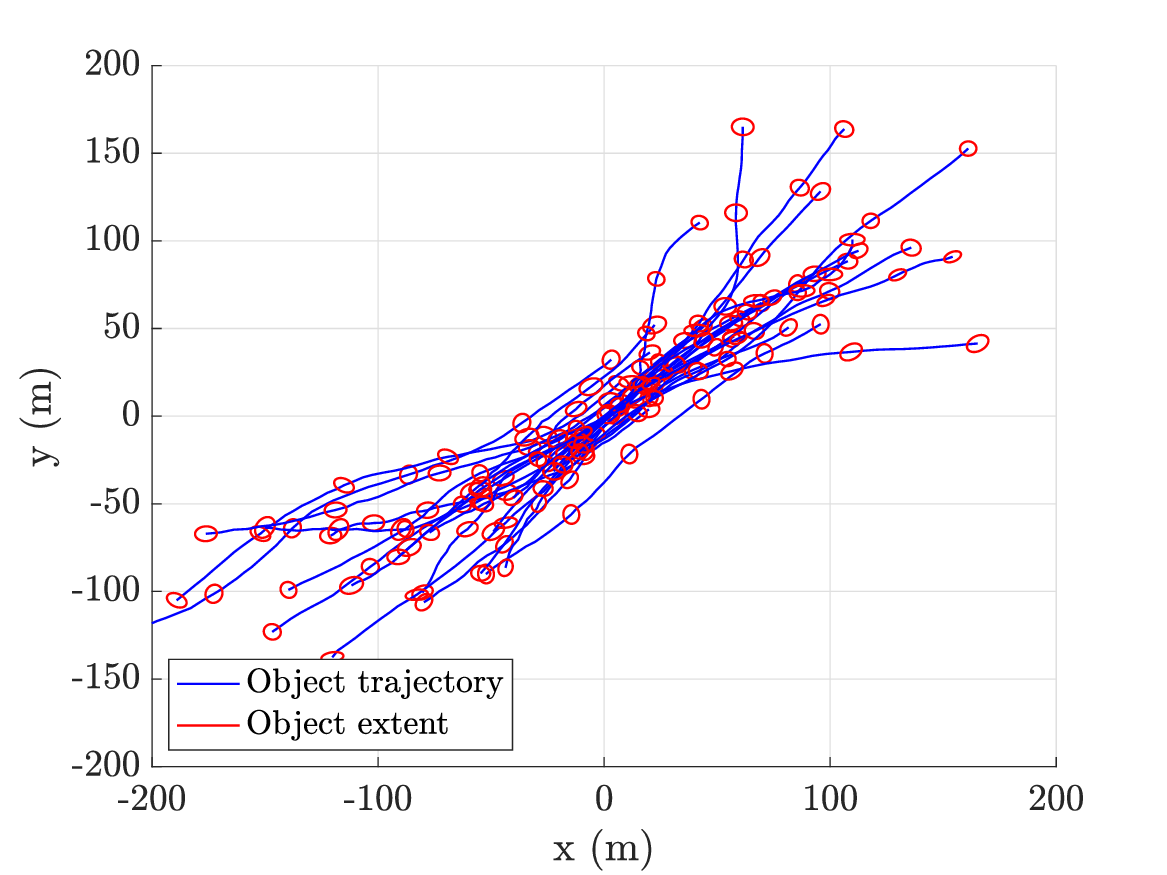}}
    \subfloat[Scenario with 8 objects]
    {\includegraphics[width = 0.67\columnwidth]{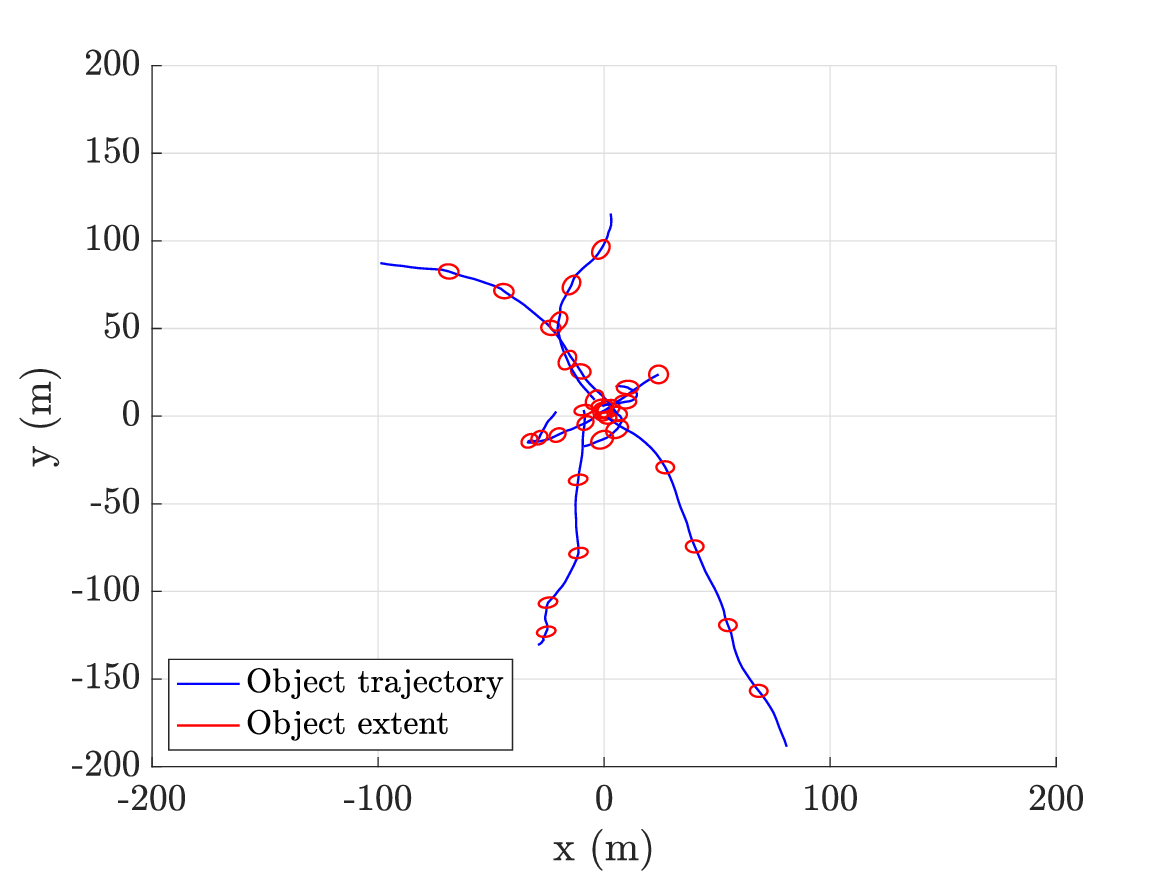}}
    \caption{True object trajectories for the simulated scenarios. The object extent is shown for every 10 time steps. In the first scenario (100 time steps), objects are born around four possible locations. In the second scenario (101 time steps), at most one object is born per time step, and objects move in proximity around the origin. In the third scenario (50 time steps), objects are all born around the same location at time step 1.}
    \label{fig:trajectory}
\end{figure*}

This section presents the results from a Monte Carlo simulation with 100 runs where the performance of eleven different extended object filtering implementations\footnote{MATLAB code is available at \url{https://github.com/yuhsuansia}.} are evaluated: \begin{enumerate}
    \item PMBM filter with track-oriented implementation \cite{pmbmextended2}.
    \item MBM filter with track-oriented implementation.
    \item TO-PMB filter with PMB update in Lemma \ref{lemma_pmbupdate} where each measurement \textit{cluster} creates a new Bernoulli component, referred to as TO-PMB-C.
    \item TO-PMB filter with PMB update in Lemma \ref{lemma_newpmbupdate} where each \textit{measurement} creates a new Bernoulli component, referred to as TO-PMB-M.
    \item PMB filter with variational MB approximation based on 2D \textit{assignment} (\ref{eq:ml_assign}), referred to as V-PMB-A.
    \item PMB filter with variational MB approximation based on \textit{linear programming} (\ref{eq:tp_lp}), referred to as V-PMB-LP.
    \item TO-MB filter, see Section \ref{tomb_filter}.
    \item MB filter with variational MB approximation based on 2D \textit{assignment}, referred to as V-MB-A.
    \item MB filter with variational MB approximation based on \textit{linear programming}, referred to as V-MB-LP.
    \item LMB filter \cite{lmbextended} with joint prediction and update \cite{vo2016efficient}.
    \item LMB filter with \textit{adaptive birth} model \cite{lmbextended} as well as joint prediction and update, referred to as LMB-AB.
\end{enumerate}
We evaluate these implementations in three different simulated scenarios, with true object trajectories and extents illustrated in Fig. \ref{fig:trajectory}. For all the scenarios, the object's survival probability is $p^S = 0.99$, the object's Poisson measurement rate is $\gamma = 10$ and its initial extent is randomly generated from an inverse-Wishart distribution with mean $\chi = 4I_2$ where $I_2$ is an identity matrix with size 2. In the first scenario, 96 randomly generated objects are born around four locations, and they appear in and disappear from the surveillance area $[-200,200]\times[-200,200]$ at different time steps. The probability of detection is $p^D = 0.9$, and the clutter is uniformly distributed in the surveillance area with Poisson rate $\lambda = 60$. In the second scenario, 26 objects first get close to each other and then separate. The probability of detection is $p^D = 0.3$ and the Poisson clutter rate is $\lambda=100$. In the third scenario, 8 objects are all born around the same location at time step $1$. The probability of detection is $p^D = 0.7$ and the Poisson clutter rate is $\lambda=30$. 

The ellipsoidal gate size in probability is $0.999$. For solving the data association problem, we first apply DBSCAN \cite{ester1996density} with different hyperparameters to obtain a set of measurement partitions. DBSCAN is also used to identify the clusters of new Bernoulli components in
PMBM and TO-PMB. We then use Murty's algorithm\footnote{The C++ implementation in the Tracker Component Library \cite{crouse2017tracker} is used.} to find the $\lceil 20\cdot w_a\rceil$ best cluster-to-object assignments for each measurement partition. The updated MBs are pruned to only contain high-weight MBs that correspond to $99\%$ of the likelihood. For filters without MB approximation, the number of updated MBs is also capped at $100$. For filters with MB birth model, Bernoulli components with probability of existence smaller than $0.001$ are pruned. For filters with Poisson birth model, Bernoulli components with probability of existence smaller than $0.1$ are recycled, and PPP mixture components with weight smaller than $0.001$ are pruned. For filters with variational MB approximation, the convergence threshold for the EM algorithm is $0.1$. For V-PMB-A and V-MB-A, the 2D assignment problem is solved using a modified Jonker-Volgenant algorithm \cite{crouse2016implementing}. For V-PMB-LP and V-MB-LP, the linear programming is addressed using Gurobi optimiser \cite{gurobi}.

The kinematic state is $\xi_k = [p_k,v_k]^T$, and it describes the object's position $p_k \in \mathbb{R}^2$ and velocity $v_k \in \mathbb{R}^2$. The symmetric positive definite random matrix $\chi_k$ is two-dimensional. Objects move with a nearly constant velocity model (\ref{eq:motionmodel}) with 
\begin{equation*}
    g(\xi_k) = \begin{bmatrix}
        I_2 & T_sI_2\\
        0_2 & I_2
    \end{bmatrix}\xi_k,\quad Q = q^2\begin{bmatrix}
        \frac{T_s^3}{3}I_2 & \frac{T_s^2}{2}I_2\\
        \frac{T_s^2}{2}I_2 & T_sI_2
    \end{bmatrix}
\end{equation*}
where $T_s = 1$ is the time interval and $q = 0.3$ is the standard acceleration deviation. The transformation matrix is $M(\xi_k) = I_2$. For the measurement model (\ref{eq:mealikelihood}), the scaling factor is $z=0.25$ and the measurement noise covariance is $R_k = 0.5I_2$.

All the filters extract state estimates by first determining the maximum a posteriori estimate of the cardinality. For filters with MB approximation, mean values of the object states are reported from the corresponding number of Bernoulli components with the highest probability of existence. For PMBM and MBM, mean values of the object states are reported from the global hypothesis with corresponing cardinality with highest weight in a similar manner, see \cite[Sec. VI.B]{pmbmpoint2} for details. 



For performance evaluation of extended object estimates with ellipsoidal extent, a comparison study in \cite{gwd} has shown that a reasonable choice is the Gaussian Wasserstein distance (GWD) metric, which is defined as \cite{givens1984class}
\begin{equation*}
    d_{\text{GW}}({\bf x},\hat{{\bf x}}) = \left\|H\left(\xi-\hat{\xi}\right)\right\|^2 + \text{Tr}\left(\chi + \hat{\chi} - 2\left(\chi^{\frac{1}{2}}\hat{\chi}\chi^{\frac{1}{2}}\right)^{\frac{1}{2}}\right),
\end{equation*}
where $\text{Tr}(X)$ is the trace of matrix $X$.
To evaluate the multi-object filtering performance, the GWD metric is integrated into the generalised optimal sub-pattern assignment (GOSPA) metric \cite{gospa} with parameters $\alpha=2$, $c=60$ and $p=1$, which allows for the decomposition of the estimation error into localisation errors and costs for missed and false objects.


\begin{table*}[t!]
    \caption{Mean GOSPA error and cycle time (seconds) per time step, averaged over 100 Monte Carlo runs.\hspace{\textwidth} Legend:Tol.--gospa error; Loc.--localisation error; Mis.--missed object error; Fal.--false object error; Cyc.--cycle time}
    \label{tab:SimulationResults}
        \centering
        \resizebox{\textwidth}{!}{
        \begin{tabular}{c | ccccc | ccccc | ccccc}
        & \multicolumn{5}{c|}{Scenario with 96 objects} & \multicolumn{5}{c}{Scenario with 26 objects} & \multicolumn{5}{c}{Scenario with 8 objects}\\
        Filter & \textsc{ Tol. } & \textsc{ Loc. } & \textsc{ Mis. } & \textsc{ Fal. } & \textsc{ Cyc. } & \textsc{ Tol. } & \textsc{ Loc. } & \textsc{ Mis. } & \textsc{ Fal. } & \textsc{ Cyc. } & \textsc{ Tol. } & \textsc{ Loc. } & \textsc{ Mis. } & \textsc{ Fal. } & \textsc{ Cyc. }\\
        \hline
        \textsc{TO-PMB-C} & 75.38 & 35.68 & 19.79 & 19.91 & 1.04 & 156.99
        & 48.12 & 85.98 & 22.88 & 0.69 & 21.62 & 5.76 & 9.77 & 6.08 & 0.18 \\
    
        \textsc{TO-PMB-M} & 75.37 & 35.69 & 19.73 & 19.95 & 0.73 & 156.71 & 48.25 & 85.86 & 22.60 & 0.57 & 21.36 & 5.80 & 9.57 & 5.99 & 0.14 \\
        
        \textsc{V-PMB-A} & 75.30 & 35.68 & 19.65 & 19.98 & 0.82 & 152.58 & 47.63 & 81.06 & 23.90 & 0.64 & 21.09 & 5.55 & 9.53 & 6.02 & 0.15 \\
    
        \textsc{V-PMB-LP} & 75.30 & 35.67 & 19.65 & 19.98 & 0.81 & 152.16 & 47.30 & 80.68 & 24.17 & 0.62 & 21.11 & 5.53 & 9.54 & 6.04 & 0.15 \\
    
        \textsc{PMBM}  & \underline{74.89} & 35.61 & 19.31 & 19.97 & 2.74 &\underline{149.02}& 57.62 & 60.00 & 31.40 & 1.81 & \underline{20.42} & 5.29 & 9.49 & 5.65 & 0.40 \\
    
        \textsc{TO-MB}  & 82.83 & 41.48 & 21.11 & 20.23 & 1.17 & 188.53 & 61.07 & 100.88 & 26.58 & 0.88 & 54.35 & 10.85 & 39.77 & 3.72 & 0.14 \\

        \textsc{V-MB-A}  & 76.89 & 36.86 & 19.03 & 21.00 & 1.47 & 168.06 & 56.61 & 80.91 & 30.54 & 1.00 & 51.22 & 10.49 & 36.72 & 4.01 & 0.14 \\

        \textsc{V-MB-LP}  & 76.71 & 36.72 & 18.98 & 21.01 & 1.35 & 167.42 & 56.01 & 81.21 & 30.20 & 0.93 & 51.19 & 10.53 & 36.64 & 4.03 & 0.14 \\

        \textsc{MBM}  & 75.56 & 35.98 & 18.67 & 20.91 & 4.64 & 169.86 & 72.27 & 54.54 & 43.04 & 3.20 & 50.17 & 10.30 & 35.58 & 4.29 & 0.37 \\

        \textsc{LMB}  & 83.03 & 40.42 & 19.19 & 23.42 & 2.07 & 249.15 & 64.81 & 98.37 & 85.97 & 0.79 & 57.48 & 11.13 & 42.73 & 3.61 & 0.23 \\

        \textsc{LMB-AB}  & 165.39 & 56.19 & 77.03 & 32.17 & 1.55 & 265.74 & 65.70 & 145.23 & 54.81 & 0.22 & 50.23 & 15.03 & 29.83 & 5.38 & 0.13 \\

        \end{tabular}}
\end{table*}

\begin{figure}[t!]
    \centering
    \includegraphics[width = \columnwidth]{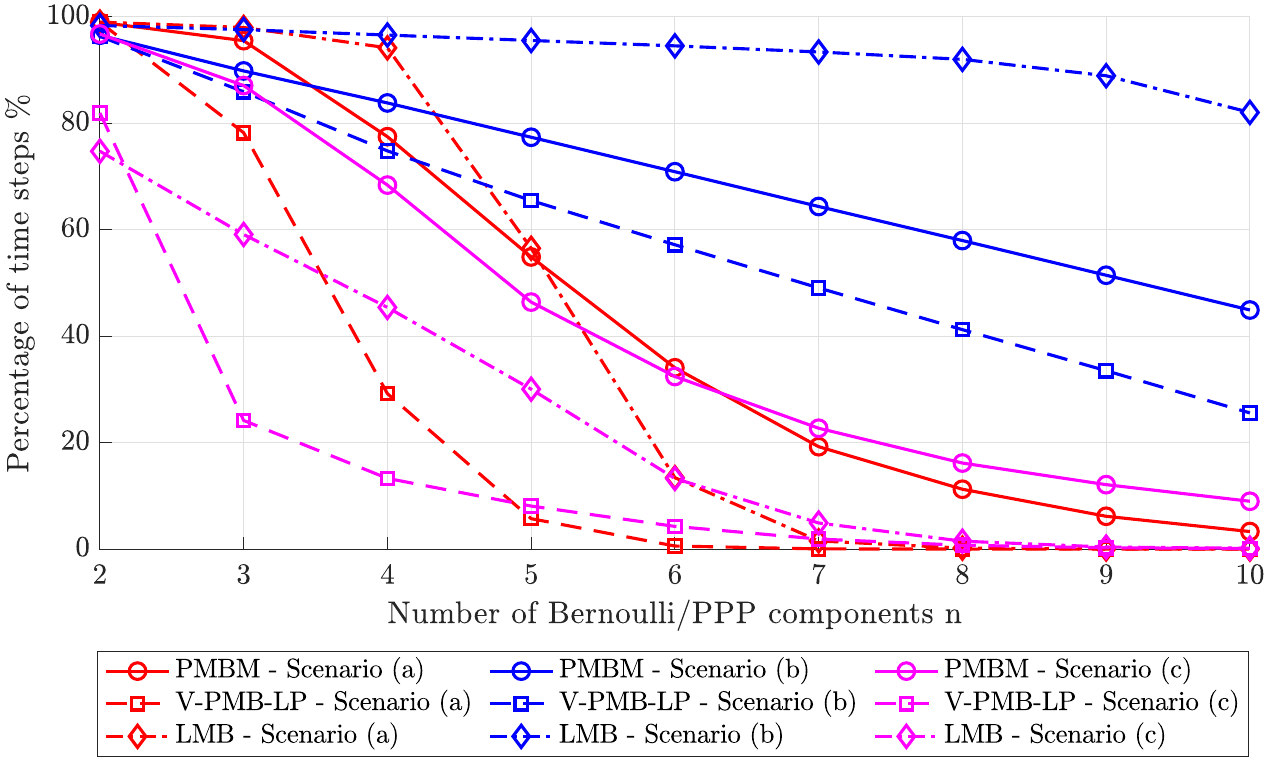}
    \caption{Percentage of time steps for which there exists at least one measurement cluster that simultaneously falls within the ellipsoidal gates of at least $n$ Bernoulli or PPP components for PMBM, V-PMB-LP and LMB in three different scenarios.}
    \label{fig:gate}
\end{figure}

For the first scenario, the Poisson birth intensity is a GGIW mixture with four components, and each of them has weight $0.25$ and Gaussian density with mean $[\pm 50,\pm 50,0,0]^T$ and covariance $\text{diag}(100,100,4,4)$ covering a potential birth area, whereas the MB birth density has four Bernoulli component, and each of them has probability of existence $0.25$ and single object GGIW density covering a potential birth area. For the second scenario, the Poisson birth intensity in this case is a single GGIW component with weight $0.25$ and Gaussian density with mean $[-100,-100,0,0]^T$ and covariance $\text{diag}(10000,10000,16,16)$ covering a large surveillance area, whereas the MB birth density only contains a single Bernoulli with probability of existence $0.25$ and the same GGIW component as it is in the Poisson birth intensity. For birth models used in the third scenario, the only difference compared to the second scenario is that the Gaussian density now has mean $[0,0,0,0]^T$ and covariance $\text{diag}(16,16,4,4)$.

For a quantitative analysis of the data association uncertainty in the three different scenarios, we also computed the percentage of time steps for which there exists at least one measurement cluster that simultaneously falls within the ellipsoidal gates of at least $n$ Bernoulli or PPP components, see Fig. \ref{fig:gate} for the results of PMBM, V-PMB-LP and LMB. As can be seen, V-PMB-LP has the lowest data association uncertainty as it maintains fewer Bernoulli components at each time step. For the three different scenarios, the data association problem is most challenging in the scenario with 26 objects where many objects move in proximity.

\begin{figure}[t!]
    \centering
    \includegraphics[width = \columnwidth]{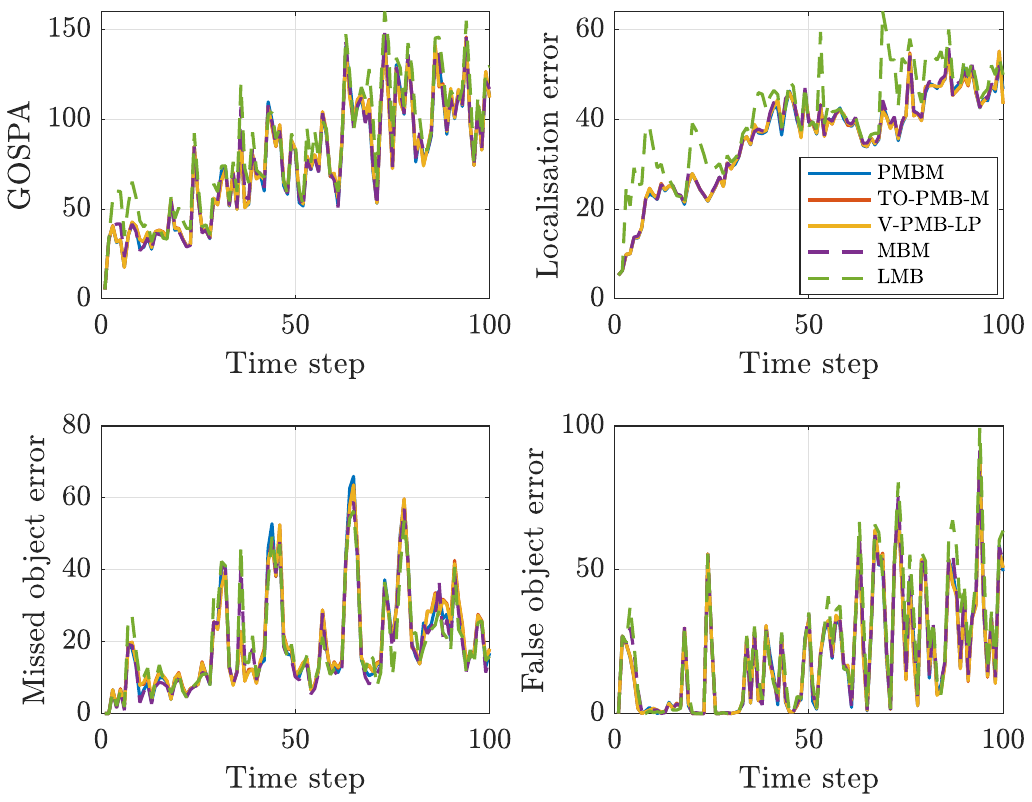}
    \caption{GOSPA error and its decomposition versus time for PMBM, TO-PMB-M, V-PMB-LP, MBM and LMB in scenario with 96 objects.}
    \label{fig:results1}
\end{figure}

\begin{figure}[t!]
    \centering
    \includegraphics[width = \columnwidth]{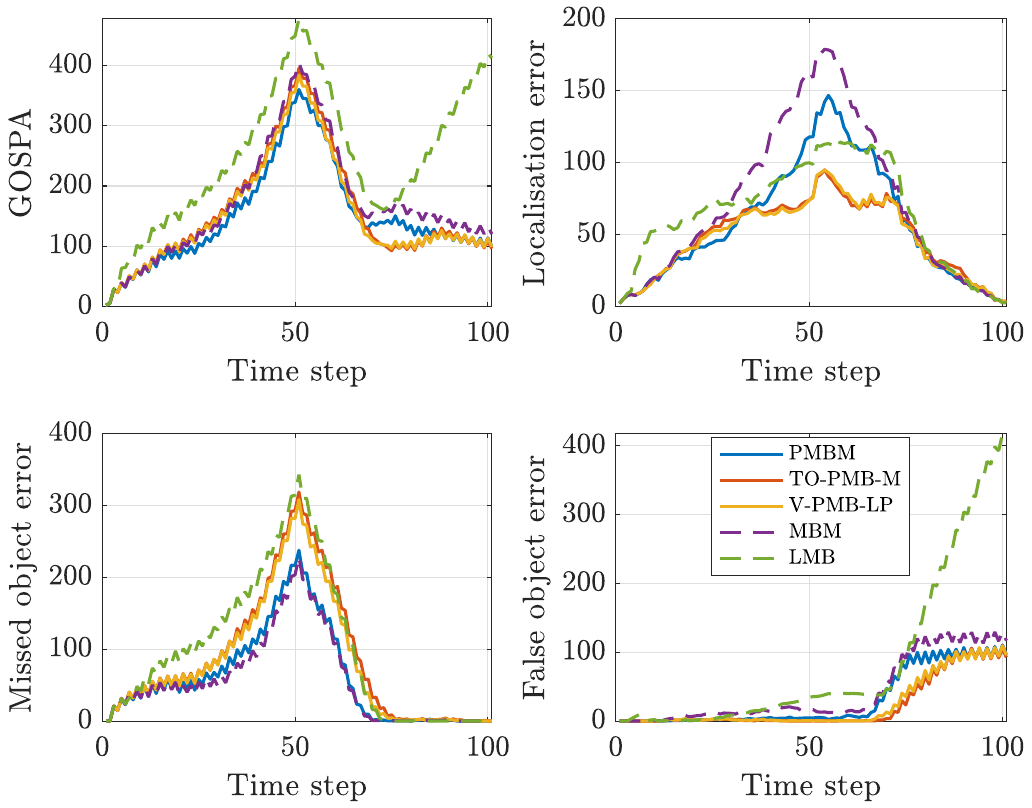}
    \caption{GOSPA error and its decomposition versus time for PMBM, TO-PMB-M, V-PMB-LP, MBM and LMB in scenario with 26 objects.}
    \label{fig:results2}
\end{figure}

\begin{figure}[t!]
    \centering
    \includegraphics[width = \columnwidth]{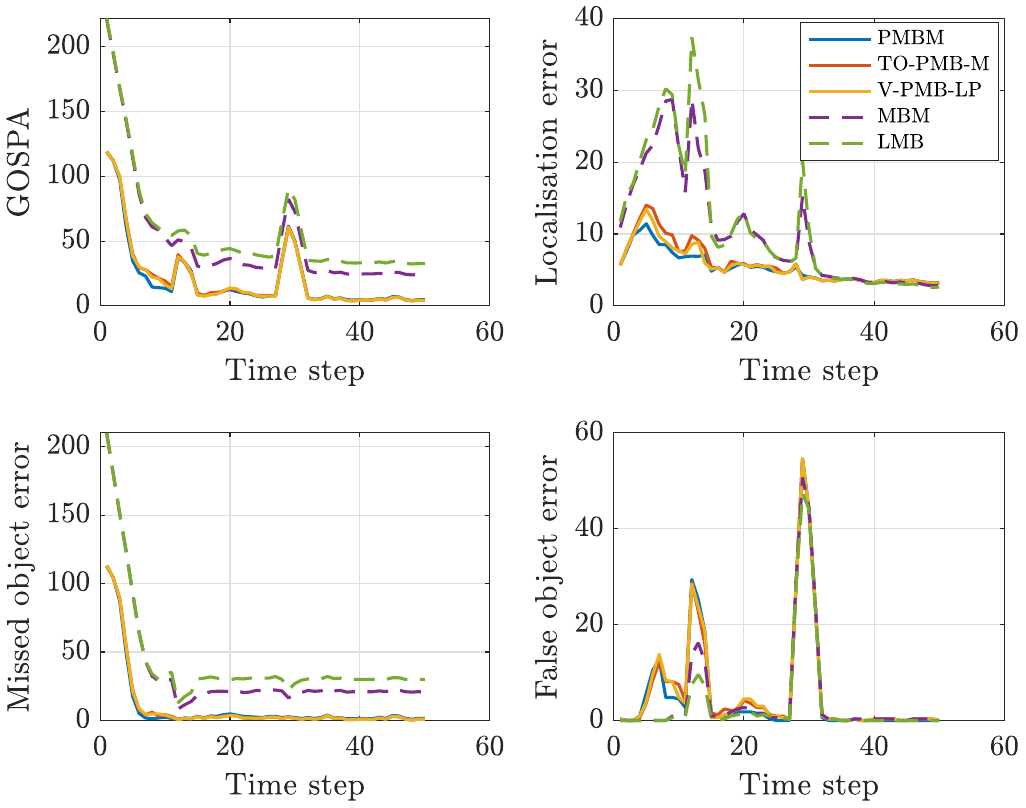}
    \caption{GOSPA error and its decomposition versus time for PMBM, TO-PMB-M, V-PMB-LP, MBM and LMB in scenario with 8 objects.}
    \label{fig:results3}
\end{figure}

The mean GOSPA error and its decomposition, along with the mean cycle time per time step\footnote{MATLAB implementation on 3 GHz 6-Core Intel Core i5.}, for the eleven different filters in two different scenarios are presented in Table \ref{tab:SimulationResults}. Also, the mean GOSPA error and its decomposition for the PMBM, MBM, TO-PMB-M, V-PMB-LP and LMB filters in three different scenarios are shown in Fig. \ref{fig:results1}, \ref{fig:results2} and \ref{fig:results3}, respectively. It can be seen that the PMBM filter has the best estimation performance for all the scenarios. LMB-AB shows the worst performance in the first scenario as it does not know the region where new objects will appear. LMB-AB outperforms LMB in the third scenario where there is a birth model mismatch. 

All filters with Poisson birth model outperform their counterparts with MB birth model in terms of both estimation error and computational time. This is mainly because the PMBM conjugate prior has a more efficient hypothesis structure than the MBM conjugate prior \cite{pmbmpoint2}. TO-PMB-C has slightly better estimation performance than TO-PMB-M, and it is also faster. The noticable difference in mean cycle time is because in TO-PMB-C many more new Bernoulli components are created in the update step, which increases the complexity of the PMB filtering recursion in subsequent time steps. 

All filters with variational MB approximation outperform their counterparts with TO-MB approximation and the difference is most noticable in the second scenario with coalescence. The multi-object density becomes strongly multimodal when objects are closely-spaced, and in this case it is less accurate to consider the TO-MB approximation. The PMB filters with variational MB approximation not only have very close estimation performance to the PMBM filter in all the three scenarios, but they are also significantly faster. For the two variants of the variational MB approximation, V-PMB-LP has similar performance compared to V-PMB-A, whereas V-MB-LP has slightly better performance than V-MB-A.

\section{Conclusions}
We have proposed two different PMB approximations of the multi-object posterior density for extended object filtering. They are both based on a new and more efficient local hypothesis representation for newly detected objects. Two PMB filters and their GGIW implementations are presented: the TO-PMB filter is based on the TO-MB approximation and the V-PMB filter is based on the variational MB approximation. The PMB filters offer a good comprise between computational complexity and accuracy. Specifically, they are generally faster than the PMBM filter with slightly worse performance. The two V-PMB filters have better performance than the TO-PMB filter, especially in scenarios with a high risk of coalescence.

\bibliographystyle{IEEEtran}
\bibliography{IEEEabrv,mybibli}

\clearpage 
\newpage
{\bfseries\huge Supplementary Materials}

\appendices

\section{}
\label{apendix:newlocalhyporepresentation}
\subsection{Proof of Theorem \ref{theorem1}}
Because the different local hypothesis representations only differ in how the newly detected objects are represented, for simplicity we only demonstrate the case with only newly detected objects, i.e., $n_{k|k-1} = 0$.


Using the local hypothesis representation in Lemma \ref{lemma_pmbupdate}, the update of a PPP prior density with a set ${\bf Z}_k = \left\{{\bf z}_k^1,\dots,{\bf z}_k^{m_k}\right\}$ of measurements is a PMBM where the MBM density describing the set ${\bf X}_k$ of newly detected objects is 
\begin{equation}
    \label{eq:mbm1}
    f_{k|k}({\bf X}_k) = \sum_{a\in{\cal A}_{k|k}}w^a_{k|k} \sum_{\biguplus_{l=1}^{2^{m_k}-1}{\bf X}^l = {\bf X}_k}\prod_{i=1}^{2^{m_k}-1}f^{i,a^i}_{k|k}\left({\bf X}^i\right)
\end{equation}
where the set 
\begin{multline}
    \label{eq:global_hypo_constraint1}
    {\cal A}_{k|k} = \left\{ \left(a^1,\dots,a^{2^{m_k}-1}\right):\right.\\\left.a^i\in\left\{1,2\right\}~\forall~i, \biguplus_{i=1}^{2^{m_k}-1}{\cal M}_k^{i,a^i} = {\cal M}_k\right\}
\end{multline}
of global hypotheses is made up of local hypotheses (\ref{eq:newtracks}). By construction, there is a one-to-one correspondence between the set of partitions of ${\cal M}_k$ and the set ${\cal A}_{k|k}$. 

Now consider a local hypothesis representation with $n_k \geq m_k$ Bernoulli components, and the MBM density describing the set ${\bf X}_k$ of newly detected objects is instead expressed as
\begin{equation}
    \label{eq:mbm2}
    f^\prime_{k|k}({\bf X}_k) = \sum_{b\in{\cal B}_{k|k}}w^b_{k|k} \sum_{\biguplus_{l=1}^{n_k}{\bf X}^l = {\bf X}_k}\prod_{i=1}^{n_k}f^{i,b^i}_{k|k}\left({\bf X}^i\right)
\end{equation}
where the set
\begin{multline}
    \label{eq:global_hypo_constraint2}
    {\cal B}_{k|k} = \left\{ \left(b^1,\dots,b^{n_{k}}\right):\right.\\\left.b^i\in\left\{1,\dots,h^i_{k|k}\right\}~\forall~i, \biguplus_{i=1}^{n_{k}}{\cal M}_k^{i,b^i} = {\cal M}_k\right\}
\end{multline}
of global hypotheses is made up of local hypotheses whose associations ${\cal M}_k^{i,j}$ satisfy (\ref{eq:constrainsnewlocal}). Note that for a local hypothesis with ${\cal M}_k^{i,j}$ its weight $w^{i,j}_{k|k}$ and density $f^{i,j}_{k|k}(\cdot)$ are computed in the same manner (cf. (\ref{eq:newtracks})) in both representations (\ref{eq:mbm1}) and (\ref{eq:mbm2}). The objective is to prove that (\ref{eq:mbm1}) and (\ref{eq:mbm2}) represent the same MBM density.

We use the following two lemmas, which are proved later in Appendix \ref{appendix_A_lemma1} and \ref{appendix_A_lemma2}, to proceed with the proof. 
\begin{lemma}
    \label{lemma:sameglo}
If global hypotheses $a$ and $b$ represent the same partition of ${\cal M}_k$, the two corresponding weights and MB densities given by $a\in{\cal A}_{k|k}$ in (\ref{eq:mbm1}) and $b\in{\cal B}_{k|k}$ in (\ref{eq:mbm2}), respectively, are identical. That is, it holds that
\begin{subequations}
    \label{eq:sameMB}
    \begin{align}
        w^a_{k|k} &= w^b_{k|k},\label{eq:sameweight}\\
        \sum_{\biguplus_{l=1}^{2^{m_k}-1}{\bf X}^l = {\bf X}_k}\prod_{i=1}^{2^{m_k}-1}f^{i,a^i}_{k|k}\left({\bf X}^i\right) &= \sum_{\biguplus_{l=1}^{n_k}{\bf X}^l = {\bf X}_k}\prod_{\iota=1}^{n_k}f^{\iota,b^\iota}_{k|k}\left({\bf X}^\iota\right)\label{eq:sameber}
    \end{align} 
\end{subequations}
where the global hypothesis weights $w^a_{k|k}$ and $w^b_{k|k}$ can be computed according to (\ref{eq:globalhypothesisweight}) using the local hypothesis weights $\left\{w^{i,a^i}_{k|k}\right\}$ and $\left\{w^{\iota,b^\iota}_{k|k}\right\}$, respectively. 
\end{lemma}


\begin{lemma}
    \label{lemma:onetoone}
    For a local hypothesis representation with $n_k \geq m_k$ Bernoulli components satisfying (\ref{eq:constrainsnewlocal}), there is a one-to-one correspondence between its corresponding set ${\cal B}_{k|k}$ of global hypotheses, given by (\ref{eq:global_hypo_constraint2}), and the set of partitions of ${\cal M}_k$.
\end{lemma}

Since there is also a one-to-one correspondence between the set of partitions of ${\cal M}_k$ and the set ${\cal A}_{k|k}$, for every $a\in {\cal A}_{k|k}$ Lemma \ref{lemma:onetoone} ensures that there is precisely one $b\in{\cal B}_{k|k}$ that corresponds to the same partition of ${\cal M}_{k}$. According to Lemma \ref{lemma:sameglo}, we can conclude that (\ref{eq:mbm1}) and (\ref{eq:mbm2}) represent the same MBM density.

This completes the proof. 

\subsection{Proof of Lemma \ref{lemma:sameglo}}
\label{appendix_A_lemma1}
The constraints (\ref{eq:global_hypo_constraint1}) and (\ref{eq:global_hypo_constraint2}), on the global hypotheses, ensure that each global hypothesis is a collection of local hypotheses, one for each Bernoulli component, and that the set of local hypotheses with non-empty ${\cal M}_k^{i,j}$ in a global hypothesis corresponds to a partition of ${\cal M}_k$. Therefore, if global hypotheses $a$ and $b$ represent the same partition of ${\cal M}_k$, their corresponding sets of local hypotheses with non-empty ${\cal M}_k^{i,j}$ are the same, i.e., 
\begin{multline*}
    \left\{ \left( {\cal M}_k^{i,a^i}, w_{k|k}^{i,a^i}, f_{k|k}^{i,a^i}(\cdot)\right) \right\}_{{\cal M}_k^{i,a^i}\neq \emptyset} \\= \left\{ \left({\cal M}_k^{\iota,b^\iota}, w_{k|k}^{\iota,b^\iota}, f_{k|k}^{\iota,b^\iota}(\cdot)\right)\right\}_{{\cal M}_k^{\iota,b^\iota}\neq \emptyset}.
\end{multline*}
    
Local hypotheses associated with empty ${\cal M}_k^{i,j}$ have local hypothesis weight $w_{k|k}^{i,j}=1$ and Bernoulli density $f_{k|k}^{i,j}(\emptyset)=1$ (cf. \ref{eq:nonexistent}). Therefore, adding these local hypothese to a global hypothesis does not change its weight and density. This means that, for global hypotheses $a\in{\cal A}_{k|k}$ and $b\in{\cal B}_{k|k}$, as long as their corresponding sets of local hypotheses with non-empty ${\cal M}_k^{i,j}$ represent the same partition of ${\cal M}_{k}$, (\ref{eq:sameMB}) holds, i.e., they have the same weight (\ref{eq:sameweight}) and MB density (\ref{eq:sameber}). 

This completes the proof. 

\subsection{Proof of Lemma \ref{lemma:onetoone}}
\label{appendix_A_lemma2}
Let us first establish that there is at least one $b\in{\cal B}_{k|k}$ for every partition of ${\cal M}_{k}$. To this end, given an arbitrary partition ${\cal P}$ of ${\cal M}_k$, we will now show that there is a corresponding $b \in {\cal B}_{k|k}$. Constraint (\ref{eq:theorem1a}) means that the local hypothesis representation contains all subsets in ${\cal P}$. This ensures that for each element of ${\cal P}$ we can find an identical set ${\cal M}_k^{i,j}$. Constraint (\ref{eq:theorem1e}) means that two different Bernoulli components cannot have local hypotheses whose ${\cal M}_k^{i,j}$ have non-empty intersection. This ensures that two elements in ${\cal P}$ can never be associated to local hypotheses of the same Bernoulli component. Finally, (\ref{eq:theorem1b}) ensures that for every Bernoulli component $i$, there is a local hypothesis with empty ${\cal M}_k^{i,j}$. To construct a global hypothesis $b\in {\cal B}_{k|k}$ that corresponds to ${\cal P}$, we can first go through the elements of ${\cal P}$, and, for each element, we find an identical set ${\cal M}_k^{i,j}$ and set $b^i = j$. At last, for each of the remaining Bernoulli components, we find its local hypothesis with empty ${\cal M}_k^{i,j}$ and set $b^i = j$. To summarise, constraints (\ref{eq:theorem1a}), (\ref{eq:theorem1b}) and (\ref{eq:theorem1e}) together ensure that there is at least one $b\in{\cal B}_{k|k}$ for every partition of ${\cal M}_{k}$.

To complete the proof, it remains to establish that no partition ${\cal P}$ is represented by two different $b\in{\cal B}_{k|k}$. Constraints (\ref{eq:theorem1c}), (\ref{eq:theorem1d}) imply that each Bernoulli component has only one local hypothesis with empty ${\cal M}_k^{i,j}$ and that there does not exist two local hypotheses with the same non-empty subset of ${\cal M}_k$. Therefore, the uniqueness of each global hypothesis $b\in{\cal B}_{k|k}$ is ensured.

This completes the proof.

\section{}
\label{appendix:bernoullimixturereduction}

In this appendix, we present the merging of a mixture of Bernoulli densities, in the sense of KLD minimisation. 

Let $f(\setX) = \sum_{h\in{\cal H}} w^{h} f^{h}(\setX)$ be a mixture of Bernoulli components, where the single object density of Bernoulli component $f^{h}(\setX)$ is from a family of distributions $\mathcal{F}$, i.e., $p^{h}(\mathbf{x})\in\mathcal{F}~\forall~h\in{\cal H}$. The objective is then to approximate the Bernoulli mixture $f(\setX)$ by a single Bernoulli distribution $\widetilde{f}(\setX)$, whose single object density is from the the same family of distributions, i.e., $\widetilde{p}(\mathbf{x})\in\mathcal{F}$. The approximate Bernoulli density $\widetilde{f}(\setX)$ that minimises the KLD $D_{\text{KL}}\left(f(\setX) \|\widetilde{f}(\setX) \right)$ has probability of existence and single object density \cite{pmbmpoint}:
\begin{subequations}
\begin{align}
    \widetilde{r} &=   \sum_{h\in\mathcal{H}} w^{h}r^{h}, \label{eq:epmr}\\
    \widetilde{p}(\sx) &= \operatornamewithlimits{argmin}_{p({\bf x})\in{\cal F}}~D_{\text{KL}} \left.\left(  \frac{\sum_{h\in\mathcal{H}} w^{h} r^{h} p^{h}(\sx)}{\sum_{h\in\mathcal{H}} w^{h}r^{h}}  \right\| p(\sx) \right). \label{eq:StateDensityKLdivMinimisation}
\end{align}
\label{eq:approximatedBernoulli}%
\end{subequations}
For distributions from the exponential family, the KLD minimisation problem \eqref{eq:StateDensityKLdivMinimisation} can be analytically solved by matching the expected sufficient statistics.


\section{}
\label{appendix:ggiw_pseudocode}
In this appendix, we present the pseudo code of factorised GGIW prediction and update with non-negligible measurement noise covariance \cite{pmbmextended2,granstrom2015gamma}, see Table \ref{tab:ggiw_prediction} and Table \ref{tab:ggiw_update}, respectively.
\begin{table}[t!]
    \begin{center}
    \caption{GGIW Prediction}
    \label{tab:ggiw_prediction}
    \begin{tabular}{p{0.95\columnwidth}}
        \toprule
        \textbf{Input:} GGIW parameter $\zeta$, motion model $g(\cdot)$, process noise covariance $Q$, transformation matrix $M(\cdot)$, time interval $T_s$, maneuvering correlation constant $\tau$, measurement rate parameter $\eta$, dimension $d$ of the extent state\\
        \textbf{Output:} Predicted GGIW parameter $\zeta_+$.\\
        \begin{equation*}
            \zeta_+ = \begin{cases}
                a_+ &= \frac{a}{\eta},\\
                b_+ &= \frac{b}{\eta},\\
                m_+ &= g(m),\\
                P_+ &= GPG^T + Q,\\
                v_+ &= 2d+2+e^{-T_s/\tau}(v-2d-2),\\
                V_+ &= e^{-T_s/\tau}M(m)VM(m)^T,
            \end{cases}
        \end{equation*}
        where $G = \nabla_{\xi}g(\xi) \mid_{\xi = m}$.\\
        \bottomrule
    \end{tabular}
    \end{center}
\end{table}

\begin{table}[t!]
    \begin{center}
    \caption{GGIW Update}
    \label{tab:ggiw_update}
    \begin{tabular}{p{0.95\columnwidth}}
        \toprule
        \textbf{Input:} GGIW parameter $\zeta_+$, set ${\bf W}$ of detections, measurement model $H$, measurement noise covariance $R$, scaling factor $z$, dimension $d$ of the extent state\\
        \textbf{Output:} Updated GGIW parameter $\zeta$ and predicted likelihood $\ell$.\\
        \begin{equation*}
            \zeta = \begin{cases}
                a &= a_+ + |{\bf W}|,\\
                b &= b_+ + 1,\\
                m &= m_+ + K\epsilon,\\
                P &= P_+ - KHP_+,\\
                v &= v_+ + |{\bf W}|,\\
                V &= V_+ + N + \hat{Z},
            \end{cases}
        \end{equation*}
        \begin{equation*}
            \ell = \pi^{\frac{-d|{\bf W}|}{2}}|{\bf W}|^{-\frac{d}{2}}\frac{|V_+|^{\frac{v_+-d-1}{2}}\Gamma_d\left(\frac{v-d-1}{2}\right)\left|\hat{X}\right|^{\frac{|{\bf W}|}{2}}\Gamma(a)(b_+)^{a_+}}{|V|^{\frac{v-d-1}{2}}\Gamma_d\left(\frac{v_+-d-1}{2}\right)\left|\hat{R}\right|^{\frac{|\bf{W}|-1}{2}}|S|^{\frac{1}{2}}\Gamma(a_+)(b)^a}
        \end{equation*}
        where
        \begin{center}
            {$\begin{aligned}  
                \bar{{\bf z}} &= \frac{1}{|{\bf W}|} \sum_{{\bf z}^i\in{\bf W}}{\bf z}^i, \\
                Z &=  \sum_{{\bf z}^i\in{\bf W}} \left({\bf z}^i - \bar{{\bf z}}\right)\left({\bf z}^i - \bar{{\bf z}}\right)^T,\\ 
                \hat{X} &= V_+(v_+-2d-2)^{-1},\\
                \epsilon &= \bar{{\bf z}} - Hm_+,\\
                \hat{R} &= z\hat{X} + R,\\
                S &= HP_+H^T + \frac{\hat{R}}{|{\bf W}|},\\
                K &= P_+H^TS^{-1},\\
                N &= \hat{X}^{1/2}\hat{S}^{-1/2}\epsilon\epsilon^T\hat{S}^{-1/2}\hat{X}^{T/2},\\
                \hat{Z} &= \hat{X}^{1/2}\hat{R}^{-1/2}Z\hat{R}^{-1/2}\hat{X}^{T/2}.\\
            \end{aligned}$}
        \end{center}\\
        \bottomrule
    \end{tabular}
    \end{center}
\end{table}

\section{}
\label{appendix:crossentropy}
In this appendix, we present the cross entropy between two Bernoulli-GGIWs. Suppose $f^h(\cdot)$ and $\widetilde{f}^i(\cdot)$ are two Bernoulli densities with the following form
\begin{subequations}
\begin{align}
    f^{h}(\mathbf{X}) &= \begin{cases}
        1 - r^{h} & \mathbf{X} = \emptyset,\\
        r^{h}\mathcal{GGIW}\left(\mathbf{x}^{h};\zeta^{h}\right) & \mathbf{X} = \{\mathbf{x}\},
    \end{cases}\\
    \widetilde{f}^{i}(\mathbf{X}) &= \begin{cases}
        1 - r^{i} & \mathbf{X} = \emptyset,\\
        r^{i}\mathcal{GGIW}\left(\mathbf{x}^{i};\zeta^{i}\right) & \mathbf{X} = \{\mathbf{x}\},
    \end{cases}
\end{align}
\end{subequations}
where $\zeta = (a,b,m,P,v,V)$ is the GGIW density parameters. 
The negative cross entropy between $f^h(\cdot)$ and $\widetilde{f}^i(\cdot)$ is
\begin{subequations}
    \begin{multline}
        \int f^{h}(\mathbf{X})\log \widetilde{f}^{i}(\mathbf{X})\delta\mathbf{X} = \left(1-r^{h}\right)\log\left(1-r^i\right) + r^{h}\log r^i \\ +r^{h}\left(\int\mathcal{N}\left(\xi;m^{h},P^{h}\right)\log\mathcal{N}\left(\xi;m^i,\hat{P}^i\right)d\xi\right. \\ +\int\mathcal{GAM}\left(\gamma;a^{h},b^{h}\right)\log\mathcal{GAM}\left(\gamma;a^i,b^i\right)d\gamma\\+ \left.\int\mathcal{IW}\left(\chi;v^{h},V^{h}\right)\log\mathcal{IW}\left(\chi;v^i,V^i\right)d\chi\right),
    \end{multline}
\begin{multline}
    \int\mathcal{N}\left(\xi;m^{h},P^{h}\right)\log\mathcal{N}\left(\xi;m^i,P^i\right)d\xi =\\ -\frac{d_\xi}{2}\log(2\pi)-\frac{1}{2}\log\left|P^i\right|\\-\frac{1}{2}\text{Tr}\left(\left(P^{h}+\left(m^{h}-m^i\right)\left(m^{h}-m^i\right)^T\right)\left(P^i\right)^{-1}\right),
    \end{multline}
    \begin{multline}
        \int\mathcal{GAM}\left(\gamma;a^{h},b^{h}\right)\log\mathcal{GAM}\left(\gamma;a^i,b^i\right)d\gamma= a^i\log b^i \\- \log\Gamma\left(a^i\right) + \left(a^i-1\right)\left(\psi_0\left(a^{h}\right)-\log b^{h}\right) - b^i\frac{a^{h}}{b^{h}},
    \end{multline}
    \begin{multline}
        \int\mathcal{IW}\left(\chi;v^{h},V^{h}\right)\log\mathcal{IW}\left(\chi;v^i,V^i\right)d\chi =\\ -\frac{d_\chi+1}{2}\log\left|\frac{V^h}{2}\right| + \frac{v^i}{2}\sum_{j=1}^{d_\chi} \psi_0\left(\frac{v^h-d_\chi-j}{2}\right) \\ -\log \Gamma_{d_\chi}\left(\frac{v^i-d_\chi-1}{2}\right) - \frac{v^h-d_\chi-1}{2}\text{Tr}\left(\left(V^h\right)^{-1}V^i\right) \\ +\frac{v^i-d_\chi-1}{2}\log\left| \left(V^h\right)^{-1}V^i \right|,
    \end{multline}
\end{subequations}
where $d_\xi$ is the dimension of $\xi$, $d_\chi$ is the dimension of $\chi$, $\Gamma(\cdot)$ is the gamma function, $\Gamma_{d_\chi}(\cdot)$ is the multivariate gamma function, and $\varphi_0(\cdot)$ is the digamma function.

\end{document}